\documentclass[11pt]{article}
\usepackage{jcappub}
\usepackage{bm}
\usepackage{color}
\usepackage{graphicx}
\def\nn{\nonumber}
\newcommand{\be}{\begin{equation}}
\newcommand{\ee}{\end{equation}}
\newcommand{\een}{\end{subequations}}
\newcommand{\ben}{\begin{subequations}}
\newcommand{\beq}{\begin{eqalignno}}
\newcommand{\eeq}{\end{eqalignno}}

\newcommand{\lsim}{\mathrel{\mathop{\kern 0pt \rlap
      {\raise.2ex\hbox{$<$}}}\lower.9ex\hbox{\kern-.190em $ \sim$}}}
\newcommand{\gsim}{\mathrel{\mathop{\kern 0pt
      \rlap{\raise.2ex\hbox{$>$}}}\lower.9ex\hbox{\kern-.190em $\sim$}}}

\newcommand{\CO}{\mathcal{O}}
\newcommand{\Ed}{E^{\prime}}

\title{Generalized spin--dependent WIMP--nucleus interactions and the
  DAMA modulation effect}

\author[a]{Stefano Scopel,}
\author[b]{Kook-Hyun Yoon}
\author[c]{Jong-Hyun Yoon}
\emailAdd{scopel@sogang.ac.kr}
\emailAdd{koreasds@naver.com}
\emailAdd{pledge200@gmail.com}
\affiliation{Department of Physics, Sogang University, Seoul, South Korea}

\abstract{Guided by non--relativistic Effective Field Theory (EFT) we
  classify the most general spin--dependent interactions between a
  fermionic Weakly Interacting Massive Particle (WIMP) and nuclei, and
  within this class of models we discuss the viability of an
  interpretation of the DAMA modulation result in terms of a signal
  from WIMP elastic scatterings using a halo--independent approach.
  We find that, although several relativistic EFT's can lead to a
  spin--dependent cross section, in some cases with an explicit,
  non-negligible dependence on the WIMP incoming velocity, three main
  scenarios can be singled out in the non--relativistic limit which
  approximately encompass them all, and that only differ by their
  dependence on the transferred momentum.  For two of them
  compatibility between DAMA and other constraints is possible for a
  WIMP mass below 30 GeV, but only for a WIMP velocity distribution in
  the halo of our Galaxy which departs from a Maxwellian. This is
  achieved by combining a suppression of the WIMP effective coupling
  to neutrons (to evade constraints from xenon and germanium
  detectors) to an explicit quadratic or quartic dependence of the
  cross section on the transferred momentum (that leads to a relative
  enhancement of the expected rate off sodium in DAMA compared to that
  off fluorine in droplet detectors and bubble chambers). For larger
  WIMP masses the same scenarios are excluded by scatterings off
  iodine in COUPP.}

\begin{document}

\maketitle

\section{Introduction}
\label{sec:introduction}

Many underground experiments are currently searching for Weakly
Interacting Massive Particles (WIMPs), which are the most popular
candidates to provide the Dark Matter (DM) which is believed to make
up 27\% of the total mass density of the Universe\cite{planck}, and
more than 90\% of the halo of our Galaxy.  One of them
(DAMA\cite{dama}) has been observing for more than 15 years a yearly
modulation effect in the low part of its energy spectrum which is
consistent with that expected due to the Earth rotation around the Sun
from the elastic scattering of WIMPs off the sodium iodide nuclei that
constitute the crystals of its scintillators.  Many experimental
collaborations using nuclear targets different from $NaI$ and various
background--subtraction techniques to look for WIMP--elastic
scattering (LUX\cite{lux}, XENON100\cite{xenon100},
XENON10\cite{xenon10}, KIMS\cite{kims,kims_modulation},
CDMS-$Ge$\cite{cdms_ge}, CDMSlite \cite{cdms_lite},
SuperCDMS\cite{super_cdms}, SIMPLE\cite{simple}, COUPP\cite{coupp},
PICASSO\cite{picasso}, PICO-2L\cite{pico2l}) have failed to observe
any anomaly so far, implying severe constraints on the most popular
WIMP scenarios used to explain the DAMA excess. This is particularly
compelling when the KIMS bound \cite{kims} (which is obtained using
$CsI$ crystals) is taken at face value, and when the WIMP mass is
large enough to rule out the possibility that the DAMA effect is
mainly due to WIMP scatterings on sodium. In that case both DAMA and
KIMS should observe the same WIMP--iodine process and no room is left
to model building to reconcile the discrepancy (although several
experimental uncertainties might still be advocated to question the
robustness of the KIMS bound\footnote{The raw background of KIMS is
  about a factor of three larger than that of DAMA, so that KIMS
  sensitivity relies on background subtraction techniques that can
  potentially introduce systematic errors. Moreover KIMS has recently
  published a new measurement of its quenching factor
  \cite{kims_quenching_2015} significantly smaller than that used for
  its published bounds\cite{kims,kims_modulation} and reducing its
  sensitivity to WIMPs scatterings at low energy.}). However, when the
WIMP mass is small enough it is possible to assume that the
contribution of WIMP-iodine scatterings in the expected event rates of
both DAMA and KIMS vanishes and that the DAMA effect is due to
scatterings on sodium. In this case the constraints on DAMA can be
potentially relaxed by considering models where the expected WIMP
scattering rate on sodium is enhanced compared to that off the nuclei
used to obtain the experimental bounds.

Moreover, the energy dependence of the WIMP--induced scattering
spectrum depends on the velocity distribution $f(\vec{v})$ of the
incoming WIMPs.  Historically, for the latter the isothermal sphere
model has been assumed, i.e. a thermalized non--relativistic gas
described by a Maxwellian distribution whose temperature
$v_{rms}\simeq$ 270 km/sec is determined from the galactic rotational
velocity by assuming equilibrium between gravitational attraction and
WIMP pressure. However, while the presence of a thermalized component
in the WIMP halo of or Galaxy is confirmed by numerical simulations
the detailed merger history of the Milky Way is not known, allowing
for the possibility of the presence of sizeable non--thermal
components for which the density, direction and speed of WIMPs are
hard to predict.

In light of the situation summarized above several new directions have
been explored in the recent past both to remove as much as possible
the dependence on specific theoretical assumptions (both of
particle--physics and astrophysical origin) from the analysis of DM
direct detection data and to extend its scope to a wider class of
models. Starting from \cite{factorization} a general strategy has been
developed \cite{mccabe_eta,gondolo_eta, noi1} to factor out the
dependence on $f(\vec{v})$ of the expected WIMP--nucleus differential
rate $dR/d E_R$ at the given recoil energy $E_R$. This approach
exploits the fact that $dR/d E_R$ depends on $f(\vec{v})$ only through
the minimal speed $v_{min}$ that the WIMP must have to deposit at
least $E_R$, i.e.:

\begin{equation}
\frac{dR}{d E_R}\propto \eta(v_{min})\equiv \int_{|\vec{v}|>v_{min}}\frac{f(\vec{v})}{|\vec{v}|}\; d^3 v.
\label{eq:eta_tilde_ex}
\end{equation}

\noindent By mapping recoil energies $E_R$ into same ranges of
$v_{min}$ the dependence on $\eta(v_{min})$ and so on $f(\vec{v})$
cancels out in the ratio of expected rates on different targets. Since
the mapping between $E_R$ and $v_{min}$ depends on the nuclear mass
the factorization of $\eta(v_{min})$ is only possible in the case of
detectors with a single nuclear target, or for which the expected rate
is dominated by scatterings on a single target. However in the
following we will extend this procedure to check the compatibility
between a candidate signal (such as that from the DAMA experiment) and
a null experiment also in the case when the latter contains different
targets.

On the particle--physics side one of the most popular scenarios for
WIMP--nucleus scattering is a spin--dependent interaction where the
WIMP particle $\chi$ is a fermion (either Dirac or Majorana) that
recoils on the target nucleus $T$ through it coupling to the spin
$\vec{S}_N$ of nucleons $N=(p,n)$:

\begin{equation} {\cal L}_{int}\propto \vec{S}_{\chi}\cdot
  \vec{S}_N=c^p\vec{S}_{\chi}\cdot \vec{S}_p+c^n\vec{S}_{\chi}\cdot
  \vec{S}_n.
\label{eq:spin_dependent}
\end{equation}

The above effective Lagrangian can arise as the low--energy limit of a
relativistic axial coupling $\bar{\chi}\gamma_5\gamma^{\mu}\chi
\bar{N}\gamma_5\gamma^{\mu}N$ (the incoming WIMP velocity in the halo
of our Galaxy is expected to be of order $10^{-3}c$).  One of the main
motivations of such scenario is the fact that the most stringent
bounds on the interpretation of the DAMA effect in terms of
WIMP--nuclei scatterings arise today from detectors using xenon
(LUX\cite{lux}, XENON100\cite{xenon100}) and germanium
(SuperCDMS\cite{super_cdms}) whose spin is mostly originated by an
unpaired neutron, while both sodium and iodine in DAMA have an
unpaired proton: as already pointed out by some
authors\cite{spin_n_suppression,spin_gelmini}, if the WIMP effective
coupling to neutrons $c^n$ is suppressed compared to that on protons
$c^p$ this class of bounds can be evaded. However this scenario is
presently constrained by droplet detectors (SIMPLE\cite{simple},
COUPP\cite{coupp}) and bubble chambers (PICASSO\cite{picasso},
PICO-2L\cite{pico2l}) which all use nuclear targets with an unpaired
proton (in particular, they all contain $^{19}F$, while SIMPLE
contains also $^{35}Cl$ and $^{37}Cl$ and COUPP uses also $^{127}I$).
As a consequence, this class of experiments have been shown to rule
out the scenario of Eq. (\ref{eq:spin_dependent}) also for $c^n\ll
c^p$ when standard assumptions are made on the WIMP local density and
velocity distribution in our Galaxy\cite{spin_gelmini,pico2l}.

The spin--dependent scenario has been further extended in
Refs. \cite{spin_freitsis,spin_arina}, where a generalization of the
coupling of Eq.(\ref{eq:spin_dependent}) has been discussed.  In this
scenario the WIMP particle couples to nucleons through a relativistic
axial coupling $\bar{\chi}\gamma_5\chi \bar{N}\gamma_5N$ through the
exchange of a light pseudoscalar, leading to a non--relativistic
coupling $(\vec{S}_{\chi}\cdot \vec{q}) (\vec{S}_N\cdot \vec{q})$
which depends on the projection of the nuclear spin component along
the transferred momentum $\vec{q}$. This scenario, which combines a
spin--dependent nuclear response function mainly sensitive to unpaired
nucleons (leading to relaxed constraints from $Ge$ and $Xe$ detectors
on DAMA when $c^n\ll c^p$) to a non--standard dependence on the
transferred momentum $\vec{q}$, has also been shown to be in tension
with bubble chambers and droplet detectors, albeit to a lesser extent,
at least for standard assumptions on the WIMP velocity distribution
\cite{spin_gelmini}.

In the present paper we wish to elaborate further on the possibility
that the non--relativistic effective coupling between WIMPS and nuclei
is due to a combination of a spin--dependent nuclear response function
and a generalized momentum dependence. In particular, we wish to
assess the viability of this kind of scenario
by departing from the Isothermal Sphere model and adopting the
previously mentioned halo--independent approach.

Our main motivation is the simple observation that models such as that
of Refs. \cite{spin_freitsis,spin_arina}, where the WIMP-nucleus cross
section depends on additional factors $q^n$ (with $n>$0) of the
transferred momentum, tend to suppress the sensitivity to the DAMA
effect of droplet detectors and bubble chambers because, due to a
combination of low energy thresholds and light target masses, a
critical fraction of the WIMP--scattering events expected in the
latter have transferred momenta below those related to the DAMA
effect.  For this reason, scenarios that combine a spin--dependent
coupling mostly to protons to additional factors $q^n$ in the cross
section are expected to have better chances to reconcile the DAMA
effect to the constraints from all other detectors. This will be
confirmed in our analysis.

In order to develop this program in a systematic way, we will rely on
the analysis of Ref.\cite{haxton}, where the most general
non--relativistic Hamiltonian {\cal H} describing the elastic
scattering of a fermionic WIMP particle off nuclei was written in
terms of the sum of all the terms invariant by Galilean
transformations. Using this approach, we will single out all the
possible spin--dependent interactions compatible to Galilean
invariance that will be the subject of our phenomenological analysis.

Our paper is organized as follows. In Section \ref{sec:eft} we use the
non--relativistic Effective Field Theory approach of Ref.\cite{haxton}
to single out the generalized spin--dependent interactions we wish to
discuss in the rest of our paper, and set our notations; in Section
\ref{sec:direct} we provide the formulas to calculate expected rates
for WIMP--nucleus scattering in our scenarios; in Section
\ref{sec:halo_independent} we summarize the main ingredients for the
halo--independent analysis; in Section \ref{sec:experiments} we
introduce a compatibility factor between DAMA and other experiments
and provide a recipe to extend it to the case of constraints involving
more than one nuclear target. The results of our quantitative analysis
are given in Section \ref{sec:results}, while Section
\ref{sec:conclusions} is devoted to our conclusions. Finally, in the
Appendices we provide some technical details of our procedure and
summarize the experimental inputs used in the analysis.

\section{Generalized spin--dependent interactions}
\label{sec:eft}

The most general WIMP--nucleus spin--dependent interactions can be
single out by making use of the non--relativistic EFT approach of
Ref.\cite{haxton}. According to that analysis the most general
Hamiltonian density for the process can be expressed in terms of a
combination of the following five Hermitian operators, which act on
the two--particle Hilbert space spanned by tensor products of WIMP and
nucleon states:

\begin{equation}
1_{\chi N} \qquad\quad i \vec{q} \qquad\quad \vec{v}^{\perp} \qquad\quad \vec{S}_{\chi} \qquad\quad \vec{S}_{N}  \, ,
\label{eq:5_operators}
\end{equation}

\noindent where $1_{\chi N}$ is the identity operator, $\vec{q}$ is
the transferred momentum, $\vec{S}_{\chi}$ and $\vec{S}_{N}$ are the
WIMP and nucleon spins, respectively, while $\vec{v}^\perp = \vec{v} +
\frac{\vec{q}}{2\mu_{\chi N}}$ (with $\mu_{\chi N}$ the WIMP--nucleon
reduced mass) is the relative transverse velocity operator satisfying
$\vec{v}^{\perp}\cdot \vec{q}=0$. Including terms that are at most
linear in $\vec{S}_N$, $\vec{S}_\chi$ and $\vec{v}^{\perp}$ only the
following 15 non-relativistic quantum mechanical operators can be
constructed out of (\ref{eq:5_operators}):

\begin{eqnarray}
  \CO_1 &=& 1_\chi 1_N; \;\;\;\; \CO_2 = (v^\perp)^2; \;\;\;\;  \CO_3 = i \vec{S}_N \cdot ({\vec{q} \over m_N} \times \vec{v}^\perp) \nn\\
  \CO_4 &=& \vec{S}_\chi \cdot \vec{S}_N;\;\;\;\; \CO_5 = i \vec{S}_\chi \cdot ({\vec{q} \over m_N} \times \vec{v}^\perp);\;\;\;\; \CO_6=
  (\vec{S}_\chi \cdot {\vec{q} \over m_N}) (\vec{S}_N \cdot {\vec{q} \over m_N}) \nn \\
  \CO_7 &=& \vec{S}_N \cdot \vec{v}^\perp;\;\;\;\;\CO_8 = \vec{S}_\chi \cdot \vec{v}^\perp;\;\;\;\;\CO_9 = i \vec{S}_\chi \cdot (\vec{S}_N \times {\vec{q} \over m_N}) \nn\\
  \CO_{10} &=& i \vec{S}_N \cdot {\vec{q} \over m_N};\;\;\;\;\CO_{11} = i \vec{S}_\chi \cdot {\vec{q} \over m_N};\;\;\;\;\CO_{12} = \vec{S}_\chi \cdot (\vec{S}_N \times \vec{v}^\perp) \nn\\
  \CO_{13} &=&i (\vec{S}_\chi \cdot \vec{v}^\perp  ) (  \vec{S}_N \cdot {\vec{q} \over m_N});\;\;\;\;\CO_{14} = i ( \vec{S}_\chi \cdot {\vec{q} \over m_N})(  \vec{S}_N \cdot \vec{v}^\perp )  \nn\\
  \CO_{15} &=& - ( \vec{S}_\chi \cdot {\vec{q} \over m_N}) ((\vec{S}_N \times \vec{v}^\perp) \cdot {\vec{q} \over m_N}),
\label{eq:ops}
\end{eqnarray}
\noindent so that the the most general Hamiltonian density describing the WIMP--nucleus interaction can be written as:
\begin{eqnarray}
{\bf\mathcal{H}}({\bf{r}})&=& \sum_{\tau=0,1} \sum_{k=1}^{15} c_k^{\tau} \mathcal{O}_{k}({\bf{r}}) \, t^{\tau} ,
\label{eq:H}
\end{eqnarray}
\noindent where $t^0=1$, $t^1=\tau_3$ denote the the $2\times2$
identity and third Pauli matrix in isospin space, respectively, and
the isoscalar and isovector (dimension -2) coupling constants $c^0_k$ and $c^{1}_k$,
are related to those to protons and neutrons $c^{p}_k$ and $c^{n}_k$
by $c^{p}_k=(c^{0}_k+c^{1}_k)/2$ and $c^{n}_k=(c^{0}_k-c^{1}_k)/2$.

The basic assumption of Ref.\cite{haxton} is that the nuclear
interaction is the sum of the interactions of the WIMPs with the
individual nucleons in the nucleus, and so neglects any effect
involving two or more nucleons. Notice that while such effects are
usually expected to be small, they can become non--negligible in
particular circumstances, such as when the leading terms is suppressed
by a cancellation between the $c^{p}_k$ and $c^{n}_k$ couplings
\cite{isospin_violation,noi2}. With this simplification the WIMP
scattering amplitude on the target nucleus $T$ can be written in the
compact form:

\begin{equation}
  \frac{1}{2 j_{\chi}+1} \frac{1}{2 j_{T}+1}|\mathcal{M}|^2=
  \frac{4\pi}{2 j_{T}+1} \sum_{\tau=0,1}\sum_{\tau^{\prime}=0,1}\sum_{k} R_k^{\tau\tau^{\prime}}\left [c^{\tau}_k,(v^{\perp}_T)^2,\frac{q^2}{m_N^2}\right ] W_{T k}^{\tau\tau^{\prime}}(y).
\label{eq:squared_amplitude}
\end{equation}

\noindent In the above expression $j_{\chi}$ and $j_{T}$ are the WIMP
and the target nucleus spins, respectively, $q=|\vec{q}|$ while the $R_k^{\tau\tau^{\prime}}$'s are WIMP response
functions (that we report for completeness in Eq.(\ref{eq:wimp_response_functions})) which depend on the couplings
$c^{\tau}_k$ as well as the transferred momentum $\vec{q}$ and on the square of the transverse velocity
in the reference frame of the center of mass of the target nucleus $(v^{\perp}_T)^2$, which can be written as:

\begin{equation}
(v^{\perp}_T)^2=v^2_T-v_{min}^2.
\label{eq:v_perp}
\end{equation}

\noindent In particular, the quantity:

\begin{equation}
v_{min}^2=\frac{q^2}{2 \mu_{T}}=\frac{m_T E_R}{2 \mu_{T}^2},
\label{eq:vmin}
\end{equation}

\noindent represents the minimal incoming WIMP speed required to
impart the nuclear recoil energy $E_R$, while $v_T\equiv|\vec{v}_T|$
is the WIMP speed in the reference frame of the nuclear center of
mass, $m_T$ the nuclear mass and $\mu_{T}$ the WIMP--nucleus reduced
mass. In equation (\ref{eq:squared_amplitude}) the
$W^{\tau\tau^{\prime}}_{T k}(q^2)$'s are nuclear response functions
and the index $k$ represents different effective nuclear operators,
which, crucially, under the assumption that the nuclear ground state
is an approximate eingenstate of $P$ and $CP$, can be at most eight:
following the notation in \cite{haxton}, $k$=$M$,
$\Phi^{\prime\prime}$, $\Phi^{\prime\prime}M$,
$\tilde{\Phi}^{\prime}$, $\Sigma^{\prime\prime}$, $\Sigma^{\prime}$,
$\Delta$,$\Delta\Sigma^{\prime}$. For the target nuclei $T$ used in
most direct detection experiments the functions
$W^{\tau\tau^{\prime}}_{T k}(q^2)$, calculated using nuclear shell
models, have been provided in Refs.\cite{haxton,catena}.

In particular, the usual spin--dependent nuclear scattering process
corresponds to ${\cal H}=\sum_{\tau=0,1}c_4^\tau {\cal O}_4t^\tau$. The corresponding form factor is
usually written in the literature in terms of the nuclear spin
structure functions  as\cite{engel_spin}:

\begin{equation}
S(q^2)=(c^0_4)^2 S_{00}(q^2)+c^0_4 c^1_4 S_{01}(q^2)+(c^1_4)^2 S_{00}(q^2).
\label{eq:spin_form_factor}
\end{equation}

\noindent At zero momentum transfer this form factor is normalized to
a combination of $j_T$ and of the expectation values $<S_{p,n}>\equiv
<T| S^z_{p,n}|T>$ of the projection $S^z$ along the $z$ axis of the
total spin operators $\vec{S}_{n,p}=\sum \vec{s}_{n,p}$ of protons and
neutrons on the nuclear state $|T>$:

\begin{equation}
S(0)=\frac{1}{\pi}\frac{(2 j_T+1)(j_T+1)}{j_T}(c^p_4 <S_p>+c^n_4 <S_n>)^2.
\label{eq:eq:spin_zero_q}
\end{equation}

\noindent In terms of the functions
$W^{\tau\tau^{\prime}}_k$ of Ref.\cite{haxton} the structure functions
$S_{\tau\tau^{\prime}}$ are given by:

\begin{eqnarray}
S_{00}(q^2)&=&W^{00}_{\Sigma^{\prime\prime}}(q^2)+W^{00}_{\Sigma^{\prime}}(q^2)\nonumber\\
S_{11}(q^2)&=&W^{11}_{\Sigma^{\prime\prime}}(q^2)+W^{11}_{\Sigma^{\prime}}(q^2)\nonumber\\
S_{01}(q^2)&=& 2 \left [W^{01}_{\Sigma^{\prime\prime}}(q^2)+W^{01}_{\Sigma^{\prime}}(q^2) \right].
\label{eq:structur_functions}
\end{eqnarray}

\noindent The above decomposition of the spin--dependent structure
functions in terms of the two different non--relativistic nuclear
response functions $\Sigma^{\prime\prime}$ and $\Sigma^{\prime}$
traces back to the decomposition:

\begin{equation}
  \vec{S}_{\chi}\cdot \vec{S}_{N}=
  (\vec{S}_{\chi}\cdot \vec{q}/q) (\vec{S}_{N}\cdot \vec{q}/q)+(\vec{S}_{\chi}\times \vec{q}/q)\cdot (\vec{S}_{N}\times \vec{q}/q).
\end{equation}

\noindent In particular, $\Sigma^{\prime\prime}$ corresponds to the
component of the nucleon spin along the direction of the transferred
momentum $\vec{q}$ while $\Sigma^{\prime}$ to that perpendicular to it
(for this reason one has
$W^{\tau\tau^{\prime}}_{\Sigma^{\prime}}(q^2)\simeq 2
W^{\tau\tau^{\prime}}_{\Sigma^{\prime\prime}}(q^2)$ when
$q^2\rightarrow 0$).

Among the $W^{\tau\tau^{\prime}}_{k}$'s, which represent the most
general nuclear response functions for WIMP--nucleus scattering, only
$W^{\tau\tau^{\prime}}_{\Sigma^{\prime\prime}}$ and
$W^{\tau\tau^{\prime}}_{\Sigma^{\prime}}$ have the property of being
suppressed for an even number of nucleons in the target, allowing to
reconcile the DAMA effect to xenon and germanium detectors by assuming
$c_i^n\rightarrow$0 (all other structure functions vanish in each
different target for a different ratio $r_i \equiv c^n_i/c^p_i$). The
correspondence between models and nuclear response functions can be
directly read off from the WIMP response functions
$R^{\tau\tau^{\prime}}_{\Sigma^{\prime\prime}}$ and
$R^{\tau\tau^{\prime}}_{\Sigma^{\prime}}$ (see
Eq.\ref{eq:wimp_response_functions}). In particular, using the
decomposition:

\be
R_k^{\tau\tau^{\prime}}=R_{0k}^{\tau\tau^{\prime}}+R_{1k}^{\tau\tau^{\prime}}\frac{(v^{\perp}_T)^2}{c^2}=R_{0k}^{\tau\tau^{\prime}}+R_{1k}^{\tau\tau^{\prime}}\frac{v_T^2-v_{min}^2}{c^2}.
\label{eq:r_decomposition}
\ee

\noindent this correspondence is summarized in Table
\ref{table:eft_summary}.
\begin{table}[t]
\begin{center}
{\begin{tabular}{@{}|c|c|c|c|c|c|@{}}
\hline
coupling  &  $R^{\tau \tau^{\prime}}_{0k}$  & $R^{\tau \tau^{\prime}}_{1k}$ & coupling  &  $R^{\tau \tau^{\prime}}_{0k}$  & $R^{\tau \tau^{\prime}}_{1k}$ \\
\hline
$1$  &   $M(q^0)$ & - & $3$  &   $\Phi^{\prime\prime}(q^4)$  & $\Sigma^{\prime}(q^2)$\\
$4$  & $\Sigma^{\prime\prime}(q^0)$,$\Sigma^{\prime}(q^0)$   & - & $5$  &   $\Delta(q^4)$  & $M(q^2)$\\
$6$  & $\Sigma^{\prime\prime}(q^4)$ & - & $7$  &  -  & $\Sigma^{\prime}(q^0)$\\
$8$  & $\Delta(q^2)$ & $M(q^0)$ & $9$  &  $\Sigma^{\prime}(q^2)$  & - \\
$10$  & $\Sigma^{\prime\prime}(q^2)$ & - & $11$  &  $M(q^2)$  & - \\
$12$  & $\Phi^{\prime\prime}(q^2)$,$\tilde{\Phi}^{\prime}(q^2)$ & $\Sigma^{\prime\prime}(q^0)$,$\Sigma^{\prime}(q^0)$ & $13$  & $\tilde{\Phi}^{\prime}(q^4)$  & $\Sigma^{\prime\prime}(q^2)$ \\
$14$  & - & $\Sigma^{\prime}(q^2)$ & $15$  & $\Phi^{\prime\prime}(q^6)$  & $\Sigma^{\prime}(q^4)$ \\
\hline
\end{tabular}}
\caption{Nuclear response functions corresponding to each coupling, for the velocity--independent and the velocity--dependent components parts of the WIMP response function, decomposed as in Eq.(\ref{eq:r_decomposition}).
  In parenthesis the power of $q$ in the WIMP response function.
  \label{table:eft_summary}}
\end{center}
\end{table}
We see from this table that the models which couple to the nucleus
either through $W^{\tau\tau^{\prime}}_{\Sigma^{\prime}}(q^2)$ or
$W^{\tau\tau^{\prime}}_{\Sigma^{\prime\prime}}(q^2)$ are ${\cal O}_i$
for $i$=3, 4, 6, 7, 9, 10, 12, 13, 14, 15. However, for models ${\cal
  O}_i$ with $i$=3, 12, 13, 15 the spin--dependent coupling is
velocity--suppressed, while the dominant velocity--independent term
couples to another nuclear response function. This implies that the
truly ``spin--dependent'' couplings are only ${\cal O}_i$ with $i$=4,
6, 7, 9, 10, 14.  In Table \ref{table:ref_nref}, which is adopted from
\cite{haxton}, we show which class of relativistic Effective Field
theories can lead to them as a non--relativistic limit.  For each
relativistic model, in the last column we provide the dependence of
the WIMP--nucleus scattering cross section on $q$, $v^{\perp}_T$,
$W^{\tau\tau^{\prime}}_{\Sigma^{\prime}}(q^2)$ and
$W^{\tau\tau^{\prime}}_{\Sigma^{\prime\prime}}(q^2)$.

Irrespective of their specific velocity
distribution, WIMP velocities are expected to be of order $10^{-3}$,
so in line 4 of Table \ref{table:ref_nref} the velocity--dependent
term is naturally suppressed. For this reason we have neglected
it. However, notice that in line 8 the coupling term
$\bar{\chi}i\sigma_{\mu\nu}\frac{q^{\nu}}{m_M}\gamma_5\chi \bar{N}
\gamma_{\mu}\gamma_5 N$ leads to a velocity--suppressed cross section
$\propto (v^{\perp}_T)^2$ without the corresponding
velocity--independent contribution. The same thing can be achieved by
appropriate linear combinations of the operator
$\bar{\chi}\gamma^{\mu}\chi \bar{N}\gamma_{\mu}\gamma^5 N$ either with
$\bar{\chi}\gamma^{\mu}\gamma_5\chi
\bar{N}i\sigma_{\mu\nu}\frac{q^{\nu}}{m_{WIMP}} N$ or
$\bar{\chi}i\sigma_{\mu\nu}\frac{q^{\nu}}{m_{WIMP}}\chi
\bar{N}\gamma^{\mu}\gamma_5 N$, as shown in lines 2 and 3 of the same
Table. In Table \ref{table:eft_summary} we ordered the models with powers of
$q^2$: actually, as we will see in the following, the scaling with
$q^2$ will be the most relevant property to interpret our numerical
results.

With the exception of line 4 and line 11, all the models of Table
\ref{table:ref_nref} correspond to one of the non--relativistic
quantum--mechanical operators of Eq.(\ref{eq:ops}). This also holds
for the model of line 4 that can be well approximated by ${\cal O}_9$.
So in the following we will refer to each of these scenarios with the
corresponding ${\cal O}_i$. On the other hand, the model of line 11 is
given by a superposition of ${\cal O}_4$ and ${\cal O}_6$ with both
contributions of the same order. In Section \ref{sec:results} we will
conventionally refer to this model as ${\cal O}_{46}$. In summary, we
conclude that the most general spin--dependent models relevant from
the point of view of the phenomenology of direct detection are seven:
${\cal O}_4$, ${\cal O}_7$, ${\cal O}_9$, ${\cal O}_{10}$, ${\cal
  O}_{14}$, ${\cal O}_6$, ${\cal O}_{46}$. We will quantitatively
discuss their phenomenology in Section \ref{sec:results}.

\begin{table}[t]
\tiny
\begin{center}
{\begin{tabular}{@{}|c|c|c|c|c|c|@{}}
\hline
 & Relativistic EFT  &  Nonrelativistic limit  & $\sum_i {\cal O}_i$ & cross section scaling\\
\hline\hline
1 & $\bar{\chi}\gamma^{\mu}\gamma^5\chi \bar{N}\gamma_{\mu}\gamma^5 N$  & $-4 \vec{S}_{\chi}\cdot\vec{S}_N$ & -4${\cal O}_{4}$ & $W^{\tau\tau^{\prime}}_{\Sigma^{\prime\prime}}(q^2)+W^{\tau\tau^{\prime}}_{\Sigma^{\prime}}(q^2)$ \\
\hline
 & $2\bar{\chi}\gamma^{\mu}\chi \bar{N}\gamma_{\mu}\gamma^5 N+$ & & & \\
2 & $+\bar{\chi}\gamma^{\mu}\gamma_5\chi \bar{N}i\sigma_{\mu\nu}\frac{q^{\nu}}{m_{WIMP}} N$  &$-4 \vec{S}_N\cdot \vec{v}^{\perp}_T$ & $-4 {\cal O}_{7}$ & $(v_T^{\perp})^2W^{\tau\tau^{\prime}}_{\Sigma^{\prime}}(q^2)$ \\
\hline
 & $2\bar{\chi}\gamma^{\mu}\chi \bar{N}\gamma_{\mu}\gamma^5 N+$ & & & \\
3 & $-\bar{\chi}i\sigma_{\mu\nu}\frac{q^{\nu}}{m_{WIMP}}\chi \bar{N}\gamma^{\mu}\gamma_5 N$  &$-4 \vec{S}_N\cdot \vec{v}^{\perp}_T$ & $-4 {\cal O}_{7}$ & $(v_T^{\perp})^2W^{\tau\tau^{\prime}}_{\Sigma^{\prime}}(q^2)$ \\
\hline

 &  &$-2 \vec{S}_N\cdot \vec{v}^{\perp}_T+$ & $-2{\cal O}_{7}+2\frac{m_N}{m_{WIMP}}{\cal O}_9$ &  \\
4 & $\bar{\chi}\gamma^{\mu}\chi \bar{N}\gamma_{\mu}\gamma^5 N$  & $+\frac{2}{m_{WIMP}}i\vec{S}_{\chi}\cdot\left (\vec{S}_N\times\vec{q} \right )$ & $\simeq2\frac{m_N}{m_{WIMP}}{\cal O}_9$  & $\simeq q^2W^{\tau\tau^{\prime}}_{\Sigma^{\prime}}(q^2)$ \\

\hline
5 & $\bar{\chi}i\sigma_{\mu\nu}\frac{q^{\nu}}{m_M}\chi \bar{N}\gamma^{\mu}\gamma_5 N$  & $4i  (\frac{\vec{q}}{m_M}\times \vec{S}_{\chi}) \cdot \vec{S}_N $ & $4\frac{m_N}{m_M}{\cal O}_{9}$ & $q^2 W^{\tau\tau^{\prime}}_{\Sigma^{\prime}}(q^2)$ \\
\hline
6 & $\bar{\chi}\gamma^{\mu}\gamma_5\chi \bar{N}i\sigma_{\mu\nu}\frac{q^{\nu}}{m_M} N$  & $4i\vec{S}_{\chi}\cdot (\frac{\vec{q}}{m_M}\times \vec{S}_N)$ & $-4\frac{m_N}{m_M}{\cal O}_{9}$ & $q^2 W^{\tau\tau^{\prime}}_{\Sigma^{\prime}}(q^2)$ \\
\hline
7 & $i\bar{\chi}\chi \bar{N}\gamma^5 N$  & $i\frac{\vec{q}}{m_N} \cdot \vec{S}_N$ & ${\cal O}_{10}$ & $q^2 W^{\tau\tau^{\prime}}_{\Sigma^{\prime\prime}}(q^2)$ \\
\hline
8 & $i\bar{\chi}i\sigma_{\mu\nu}\frac{q^{\nu}}{m_M}\gamma_5\chi \bar{N} \gamma_{\mu}\gamma_5 N$  & $-4i (\frac{\vec{q}}{m_N}\cdot \vec{S}_{\chi})(\vec{v}^{\perp}_T\cdot \vec{S}_N) $ & $-4\frac{m_N}{m_M}{\cal O}_{14}$ & $(v^{\perp}_T)^2 q^2 W^{\tau\tau^{\prime}}_{\Sigma^{\prime}}(q^2)$ \\
\hline
9 & $\bar{\chi}\gamma_5\chi \bar{N}\gamma^5 N$  & $-\frac{\vec{q}}{m_{WIMP}} \cdot \vec{S}_{\chi}\frac{\vec{q}}{m_N} \cdot \vec{S}_N$ & $-\frac{m_N}{m_{WIMP}}{\cal O}_{6}$ & $q^4 W^{\tau\tau^{\prime}}_{\Sigma^{\prime\prime}}(q^2)$ \\
\hline
10 & $\bar{\chi}i\sigma^{\mu\alpha}\frac{q_{\alpha}}{m_M}\gamma_5\chi \bar{N}i\sigma_{\mu\beta}\frac{q^{\beta}}{m_M}\gamma_5 N$  &$4\frac{\vec{q}}{m_M}\cdot\vec{S}_{\chi} \frac{\vec{q}}{m_M}\cdot\vec{S}_N $ & $4\frac{m_N^2}{m^2_M} {\cal O}_6$ & $q^4 W^{\tau\tau^{\prime}}_{\Sigma^{\prime\prime}}(q^2)$  \\
\hline
11 & $\bar{\chi}i\sigma^{\mu\nu}\frac{q_{\nu}}{m_M}\chi \bar{N}i\sigma_{\mu\alpha}\frac{q^{\alpha}}{m_M} N$  & $4\left ( \frac{\vec{q}}{m_M}\times \vec{S}_{\chi}\right )\cdot\left ( \frac{\vec{q}}{m_M}\times \vec{S}_N\right )$ & $4\left(\frac{q^2}{m_M^2}{\cal O}_4-\frac{m_N^2}{m_M^2}{\cal O}_6 \right )$ & $q^4 W^{\tau\tau^{\prime}}_{\Sigma^{\prime}}(q^2)$ \\
\hline
\end{tabular}}
\caption{Relativistic Effective Field Theories for a Dark Matter fermionic WIMP $\chi$ having as a low--energy limit a generalized spin--dependent $\chi$--nucleus elastic scattering.
In all these scenarios the scattering cross section vanishes for all nuclei with an odd number of protons if $c^p_i$=0 and for all nuclei with an odd number of neutrons if $c^n_i$=0.
Some of the interaction terms in the second column contain an arbitrary scale $m_M$ to ensure correct dimensionality.
\label{table:ref_nref}}
\end{center}
\end{table}

\section{WIMP direct detection rate}
\label{sec:direct}

For a given recoil energy imparted to the target the differential rate
for the WIMP--nucleus scattering process is given by:

\be
\frac{d R}{d E_R}=\sum_T N_T\frac{\rho_{WIMP}}{m_{WIMP}}\int_{v_{min}}d^3 v_T f(\vec{v}_T) v_T \frac{d\sigma_T}{d E_R},
\label{eq:dr_de}
\ee

\noindent where $\rho_{WIMP}$ is the local WIMP mass density in the
neighborhood of the Sun, $N_T$ the number of the nuclear targets of
species $T$ in the detector (the sum over $T$ applies in the case of
more than one nuclear isotope), $f(\vec{v}_T)$ the WIMP velocity
distribution, while

\be
\frac{d\sigma_T}{d E_R}=\frac{2 m_T}{4\pi v_T^2}\left [ \frac{1}{2 j_{\chi}+1} \frac{1}{2 j_{T}+1}|\mathcal{M}_T|^2 \right ],
\label{eq:dsigma_de}
\ee

\noindent with the squared amplitude in parenthesis given explicitly
in Eq.(\ref{eq:squared_amplitude}). Using Eqs
(\ref{eq:r_decomposition},\ref{eq:dr_de},\ref{eq:dsigma_de}) one gets:

\be
\frac{d R}{d E_R}= 2 \sum_T N_T \frac{m_T}{\sigma_{ref}}
\sum_{k\tau\tau^{\prime}} W_{T k}^{\tau\tau^{\prime}}[y(E_R)]
\left [ R_{0k}^{\tau\tau^{\prime}} \tilde{\eta}(v_{min})+R_{1k}^{\tau\tau^{\prime}} \tilde{\xi}(v_{min}) \right ],
\label{eq:dr_de_eta_xi}
\ee

\noindent where:

\begin{eqnarray}
\tilde{\eta}(v_{min})&=&\frac{\rho_{WIMP}\sigma_{ref}}{m_{WIMP}}\int_{v_{min}} d^3 \vec{v}_T \frac{f(\vec{v}_T)}{v_T} \label{eq:eta}
\nonumber\\
\tilde{\xi}(v_{min})&=&\frac{\rho_{WIMP}\sigma_{ref}}{m_{WIMP}}\int_{v_{min}} d^3 \vec{v}_T \frac{f(\vec{v}_T)}{v_T}
\left (v^{\perp}_T \right)^2=\int_{v_{min}} d^3 \vec{v}_T \frac{f(\vec{v}_T)}{v_T}\left (v^2_T-v_{min}^2 \right).
\label{eq:xi}
\end{eqnarray}

\noindent In Eq.(\ref{eq:dr_de_eta_xi}) we have factorized a reference
cross section $\sigma_{ref}$ which in the case of no momentum
dependence of the scattering amplitude can be identified with the
large--distance total cross section for a point--like nucleus. However
in the case of scattering amplitudes which vanish with $q^2$ the
quantity $\sigma_{ref}$ is just a conventional factor with dimension -2
which must be the same for all nuclei in order to cancel out
in the comparison of the expected rates in different detectors. For
all the interactions listed in Table \ref{table:ref_nref} only one of
the couplings $c^{p,n}_k$ is different from zero. In the following we
choose to factorize for each case the corresponding conventional cross
section $\sigma_{ref}=(c_k^p)^2 \mu^2_{\chi N}/\pi$.

Eq. (\ref{eq:dr_de_eta_xi}) implies that, among the models summarized
in Table \ref{table:ref_nref}, those in columns 2, 3 and 8 have a
direct detection rate proportional to $\tilde{\xi}(v_{min})$, while
all the others scale as $\tilde{\eta}(v_{min})$. In both cases the
same function can be factorized in all nuclei, allowing to adopt the
halo--independent procedure outlined in the next Section.

\section{Halo--independent analysis}
\label{sec:halo_independent}

We summarize in this Section the main formulas used in our subsequent
analysis to factorize the halo--dependence from direct--detection
data.

In a realistic experiment the recoil energy $E_R$ is obtained by
measuring a related detected energy $E^{\prime}$ obtained by
calibrating the detector with mono--energetic photons with known
energy. However the detector response to photons can be significantly
different compared to the same quantity for nuclear recoils.  For a
given calibrating photon energy the mean measured value of
$E^{\prime}$ is usually referred to as the electron--equivalent energy
$E_{ee}$ and measured in keVee. On the other hand $E_R$ (that
represents the signal that would be measured if the same amount of
energy were deposited by a nuclear recoil instead of a photon) is
measured in keVnr. The two quantities are related by a quenching
factor $Q$ according to $E_{ee}=Q(E_R) E_R$.  Moreover the measured
$E^{\prime}$ is smeared out compared to $E_{ee}$ by the energy
resolution (a Gaussian smearing
$Gauss(E_{ee}|E^{\prime},\sigma_{rms}(E^{\prime}))\equiv
1/(\sigma_{rms}\sqrt{2\pi})exp[-(E^{\prime}-E_{ee})^2/(2\sigma_{rms}^2)]$
with standard deviation $\sigma_{rms}(E^{\prime})$ related to the Full
Width Half Maximum (FWHM) of the calibration peaks at $E^{\prime}$ by
$FHWM=2.35 \sigma_{rms}$ is usually assumed) and experimental count
rates depend also on the counting efficiency or cut acceptance
$\epsilon(E^{\prime})$.  Overall, the expected differential event rate
is given by:

\begin{eqnarray}
  \frac{dR}{d E^{\prime}}&=&\epsilon(E^{\prime})\int_0^{\infty}d E_{ee} Gauss(E_{ee}|E^{\prime},\sigma_{rms}(E^{\prime}))\frac{1}{Q(E_R)} \frac{d R}{d E_R},
\label{eq:rate_folding}
\end{eqnarray}

\noindent with $d R/d E_R$ given by Eq.(\ref{eq:dr_de}). Then the number of events in a given
interval of the detected energy $E^{\prime}$ is simply:

\begin{eqnarray}
R_{[E_1^{\prime},E_2^{\prime}]}=\int_{E_1^{\prime}}^{E_2^{\prime}}  \frac{dR}{d E^{\prime}}.
\label{eq:exp_event_number}
\end{eqnarray}

\noindent Combining
Eqs.(\ref{eq:rate_folding},\ref{eq:exp_event_number}) with
Eq. (\ref{eq:dr_de_eta_xi}) and changing
variable from $E_R$ to $v_{min}$ the expected rate can be cast in the
compact form:

\be R_{[\Ed_1, \Ed_2]}=\frac{\rho_{WIMP}}{m_{WIMP}}\int_{0}^{\infty}
d\,v_{min} \tilde{\eta}(v_{min}) {\cal R}^0_{[\Ed_1,\Ed_2]}(v_{min}),
\label{eq:r_compact_eta}
\ee

\noindent for models without velocity suppression and:

\be R_{[\Ed_1, \Ed_2]}=\frac{\rho_{WIMP}}{m_{WIMP}}\int_{0}^{\infty}
d\,v_{min} \tilde{\xi}(v_{min}) {\cal R}^1_{[\Ed_1,\Ed_2]}(v_{min}),
\label{eq:r_compact_xi}
\ee

\noindent for those with velocity suppression, with:

\begin{eqnarray}
{\cal R}_{[\Ed_1,\Ed_2]}^m&\equiv&
\int_0^{\infty}d v_{min} \sum_T 8 N_T \mu_{T}^2\frac{1}{\sigma_{ref}}\sum_{k}\sum_{\tau\tau^{\prime}}R^{\tau\tau^{\prime}}_{m k}\left[q^2(v_{min}) \right]
\frac{W^{\tau\tau^{\prime}}_{T k} \left [y(v_{min}) \right ]}{2 j_T+1}
\frac{1}{Q[E_R(v_{min})]}\times \nonumber\\
&&\int_{\Ed_1}^{\Ed_2}
d \Ed \, \epsilon(\Ed) Gauss(E_{ee}[E_R(v_{min})]|E^{\prime},\sigma_{rms}(E^{\prime})),\;\;\;\;m=0,1.
\label{eq:core}
\end{eqnarray}

In the present analysis we will assume that the yearly modulation
effect observed by the DAMA experiment is due to the time--dependence
of $\tilde{\eta}$ or $\tilde{\xi}$ due to the rotation of the Earth
around the Sun. So we also introduce the modulated amplitudes:

\begin{eqnarray}
\tilde{\eta}_1(v_{min})&\equiv&\tilde{\eta}(v_{min},t=t_{max})-\tilde{\eta}(v_{min},t=t_{min}),\nonumber\\
\tilde{\xi}_1(v_{min})&\equiv&\tilde{\xi}(v_{min},t=t_{max})-\tilde{\xi}(v_{min},t=t_{min}),
\label{eq:eta_modulation}
\end{eqnarray}

\noindent with $t_{max}$ and $t_{min}$ the times of the year
corresponding to the maximum and to the minimum of the signal.  The
modulated amplitudes measured by DAMA that we will use in Section
\ref{sec:experiments} to get estimations of $\tilde{\eta}_1$ and
$\tilde{\xi}_1$ were obtained by fitting the data with a time
dependence parametrized with the functional form:

\begin{equation}
R_{[E_1^{\prime},E_2^{\prime}]}=R^0_{[E_1^{\prime},E_2^{\prime}]}+R^1_{[E_1^{\prime},E_2^{\prime}]}\cos[\omega (t-t_{max})],
\label{eq:cos_dependence}
\end{equation}

\noindent which is expected when the isothermal sphere model is
adopted for the WIMP speed distribution. Moreover, according to the
DAMA analysis, the modulation phase is in (rough) agreement to what is
expected for the isothermal sphere model, i.e. $t_{max}\simeq$ 2 June
(and, consequently, $t_{min}\simeq$ 2 December). However, a truly {\it
  halo independent} analysis would require to reanalyze the
experimental data allowing for functional forms of general nature
beyond Eq.(\ref{eq:cos_dependence})\footnote{For speed distributions
  beyond the Isothermal Sphere the time dependence may differ from
  that of Eq.(\ref{eq:cos_dependence}).  For instance, it has been
  shown that in anisotropic extensions of the isothermal sphere the
  phase and even the period may vary\cite{nic_ste}.}. Nevertheless, in
the following to estimate the $\tilde{\eta}_1$, $\tilde{\xi}_1$
functions we will make directly use of the published DAMA modulation
amplitudes, implicitly assuming that indeed those estimations do not
differ significantly from what one would obtain using a more general
form to analyze the data. With this important proviso, in the present
analysis we wish to make the smallest possible number of assumptions
on the functions $\tilde{\eta}$, $\tilde{\xi}$. They turn out to be
the same, and in the case of $\tilde{\eta}$ they reduce to:

\begin{eqnarray}
&\tilde{\eta}(v_{min,2})\le\tilde{\eta}(v_{min,1})  & \mbox{if $v_{min,2}> v_{min,1}$},\nonumber\\
& \tilde{\eta}_1\le\tilde{\eta}  & \mbox{at the same $v_{min}$},\nonumber\\
& \tilde{\eta}(v_{min} \ge v_{esc})=0. &
\label{eq:eta_conditions}
\end{eqnarray}

The first condition descends from the definitions (\ref{eq:eta}), that
implies that both $\tilde{\eta}(v_{min})$ and $\tilde{\xi}(v_{min})$
are decreasing functions of $v_{min}$. The second is an obvious
consequence of the fact that $\tilde{\eta}_1$, $\tilde{\xi}_1$ are the
modulated parts of $\tilde{\eta}$ and $\tilde{\xi}$. The last
condition reflects the requirement that the WIMPs are gravitationally
bound to our Galaxy. In the following we will assume that the WIMP
halo is at rest in the Galactic rest frame and we will adopt as the
maximal velocity of WIMPs $v_{esc}$=782 km/s in the reference frame of
the laboratory, by combining the reference value of the escape
velocity $v_{esc}^{Galaxy}$=550 km/s in the galactic rest frame with
the velocity $v_0$=232 km/s of the Solar system with respect to the
WIMP halo.

So, given an experiment with detected count rate $N_{exp}$ in the
energy interval $E_1^{\prime}<E^{\prime}<E_2^{\prime}$ the
combination:

\begin{equation}
  <\tilde{\eta}>=\frac{\int_{0}^{\infty} d v_{vmin} \tilde{\eta}(v_{min}) {\cal R}_{[E_1^{\prime},E_2^{\prime}]}(v_{min})}
  {\int_{0}^{\infty} d v_{min} {\cal R}_{[E_1^{\prime},E_2^{\prime}]}(v_{min})}=\frac{N_{exp}}{\int_{0}^{\infty} d v_{min} {\cal R}_{[E_1^{\prime},E_2^{\prime}]}(v_{min})},
\label{eq:eta_bar_vt}
\end{equation}

\noindent can be interpreted as an average of the function
$\tilde{\eta}(v_{min})$ in an interval $v_{min,1}<v_{min}<v_{min,2}$,
with an analogous definition for the average of the function
$\tilde{\xi}(v_{min})$. The $v_{min}$ interval is defined as the one
where the response function ${\cal R}$ is ``sizeably'' different from
zero (we will conventionally take the interval
$v_{min}[E_R(E_{ee,1})]<v_{min}<v_{min}[E_R(E_{ee,2})]$ with
$E_{ee,1}=E^{\prime}_1-\sigma_{rms}(E^{\prime}_1)$,
$E_{ee,2}=E^{\prime}_2+\sigma_{rms}(E^{\prime}_2)$, i.e. the
$E^{\prime}$ interval enlarged by the energy resolution).

\section{Constraints from direct detection experiments and
  compatibility factor}
\label{sec:experiments}
In this Section we will give explicit formulas for the case of a
velocity--independent cross section where the expected rate is given
by Eq.(\ref{eq:r_compact_eta}) and depends on $\tilde{\eta}(v_{min})$.
Clearly, the velocity--dependent case
of Eq.(\ref{eq:r_compact_xi}) with dependence on $\tilde{\xi}(v_{min})$ is analogous.

Following the halo--independent procedure outlined in Section
\ref{sec:halo_independent} it is straightforward, for a given choice
of the DM parameters, to obtain estimations
$\bar{\tilde{\eta}}_{1,i}^{DAMA-Na}$ of the modulation--amplitude
function $\tilde{\eta}_1(v_{min})$ averaged in appropriately chosen
$v_{min,i}$ intervals mapped from the DAMA experimental annual
modulation amplitudes. Using the condition
$\tilde{\eta}_1(v_{min})\le\tilde{\eta}(v_{min})$ this allows to get
lower bounds on the $\tilde{\eta}(v_{min})$ function, which can be
compared to upper bounds $\bar{\tilde{\eta}}_{j,lim}$ on the same
quantity derived from the data of the experiments that have reported
null results. Among the latter, the KIMS experiment uses iodine, one
of the same targets of DAMA ($NaI$ for DAMA vs. $CsI$ for KIMS). KIMS
has reported an upper bound on the unmodulated WIMP rate in the
low-part of its measured spectrum that is about a factor of two
smaller than the modulated amplitude measured by DAMA in the same
energy interval\cite{kims} . This implies that, as far as the WIMP
rate in both experiments is dominated by scatterings on iodine, the
two results cannot be reconciled by changing either the halo function
or the scaling law of the cross section. In this case the only way to
reconcile the two conflicting results is to look deeper in the
possible sources of systematic errors, including the many
uncertainties connected to quenching factors, atomic form factors,
background cuts efficiencies, etc.  Moreover, as already pointed out,
as far as DAMA is concerned the halo--function factorization is only
possible if WIMPs scatter predominantly on a single target
nucleus. For both reasons in the following we will restrict our
analysis to a range of the WIMP mass for which the DAMA effect can be
consistently explained with scatterings on sodium targets only. In the
popular case of an Isothermal Sphere for the velocity distribution
this is achieved for $m_{WIMP}\lsim$ 20 GeV. However this range for
$m_{WIMP}$ can be larger if the WIMP velocity distribution is
different from a Maxwellian and if the only assumptions on the halo
function are those of Eq.(\ref{eq:eta_conditions}). In particular it
can be maximally extended by assuming that the function $\tilde{\eta}$
is the minimal one compatible to the DAMA signal.  Assuming that the
latter corresponds for sodium to the $v_{min}$ range $v_1^{DAMA,Na}\le
v_{min}\le v_2^{DAMA,Na}$ and for iodine to the range $v_1^{DAMA,I}\le
v_{min}\le v_2^{DAMA,I}$, the most conservative assumption on
$\tilde{\eta}$ is that it vanishes for $v_{min}>v_2^{DAMA,Na}$ (in
particular, this is consistent with the minimal requirement that the
halo function $\eta$ is a decreasing function $v_{min}$.). Then, as
long as $v_1^{DAMA,I}> v_2^{DAMA,Na}$, one can conservatively assume
that scatterings on iodine are below the energy threshold of DAMA and
do not contribute to the modulation effect.  Using Eq.(\ref{eq:vmin})
this corresponds to $m_{WIMP}\lsim 60$ GeV, assuming the energy range
2 keVee $\le E^{\prime}\le$ 4 keVee for the DAMA modulation region,
enlarged by an energy resolution $\simeq 0.448\times
\sqrt{E^{\prime}}$ keVee \cite{dama}.  In the same range of $m_{WIMP}$
also scatterings on iodine and caesium are below the threshold of 3
keVee in KIMS (assuming the latest measurement of the KIMS quenching
factor \cite{kims_quenching_2015}, which is similar to that of
DAMA). As a consequence of this in the following we will restrict our
analysis to $m_{WIMP}\lsim$ 60 GeV, assume only scatterings on sodium
in DAMA and neglect the KIMS constraint.

Quantitatively, for a given choice of the WIMP mass $m_{WIMP}$, of the
interaction term from Table \ref{table:ref_nref} and of the
corresponding ratio $r_i$, the compatibility between DAMA and all the
other results can be assessed introducing the following compatibility
ratio\cite{noi2}:

\begin{equation}
  {\cal D}(m_{WIMP},r_i) \equiv \max_{i\in \mbox{signal}}
\left (\frac{\bar{\tilde{\eta}}_i^{DAMA-Na}+\sigma_i}{\min_{j\le i}\bar{\tilde{\eta}}_{j,lim}} \right ),
\label{eq:compatibility_factor_eta_i}
\end{equation}

\noindent where $\sigma_i$ represents the standard deviation on
$\bar{\tilde{\eta}}_i^{DAMA-Na}$ as estimated from the data, $i\in
\mbox{signal}$ means that the maximum of the ratio in parenthesis is
for $v_{min,i}$ corresponding to the DAMA excess, while, due to the
fact that the function $\tilde{\eta}$ is non--decreasing in all
velocity bins $v_{min,i}$, the denominator contains the most
constraining bound on $\tilde{\eta}$ for $v_{min,j}\le v_{min,i}$.
The latter minimum includes all available bounds from scintillators,
ionizators and calorimeters (see Appendix \ref{app:exp} for a summary
of the experimental inputs used in our analysis).  Specifically,
compatibility between DAMA and the constraints included in the
calculation of Eq.(\ref{eq:compatibility_factor_eta_i}) is ensured if
${\cal D}<1$.  Notice that, by combining different $v_{min,i}$ bins,
the above definition takes into account the most general momentum
dependencies predicted by the various effective interaction terms of
Table \ref{table:ref_nref}.

The procedure outlined above cannot include bubble chambers and
droplet detectors which are only sensitive to the energy threshold,
such as SIMPLE\cite{simple}(using $C_2 Cl F_5$), COUPP\cite{coupp}
(using $C F_3 I$), PICASSO\cite{picasso}(using $C_4 F_{10}$) or
PICO-2L\cite{pico2l} (using $C_3 F_8$), since in this case it is not
possible to map the corresponding bounds to arbitrary velocity
bins. Notice that these experiments contain all proton--odd elements
(fluorine, chlorine and iodine), so they are particularly relevant for
our analysis since they are complementary to germanium and xenon
detectors which are neutron--odd.  Moreover, another complication in
using the compatibility factor defined in
Eq.(\ref{eq:compatibility_factor_eta_i}) to assess agreement between
the DAMA excess and this class of experiments is that they all contain
different nuclear targets, so that it is in general not possible to
factorize the $\tilde{\eta}$ function and it is not trivial (as for
instance for the case of a coherent interaction, where the cross
section scales with the square of the atomic mass number) to single
out particular kinematic situations where one of the targets
dominates, this also because of the momentum--dependence of some of
the interactions. So, in order to impose the bounds from this class of
experiments in a halo--independent way, we need to generalize the
compatibility factor definition of
Eq.(\ref{eq:compatibility_factor_eta_i}). In order to do so, we
proceed in the following way: i) we use the experimental DAMA
modulation--amplitudes to get a conservative piecewise estimation
$\tilde{\eta}^{est}_1(v_{min})$ of the minimal $\tilde{\eta}_1$
modulated halo function compatible to the signal; ii) we obtain the
corresponding estimation of the unmodulated part
$\tilde{\eta}^{est}(v_{min})$ by requiring that it is a decreasing
function of $v_{min}$ with $\tilde{\eta}^{est}(v_{min})\ge
\tilde{\eta}^{est}_1(v_{min})$; iii) in compliance with
(\ref{eq:eta_conditions}) and with the goal of obtaining a
conservative bound, we require that the function $\tilde{\eta}$ is the
{\it minimal} one able to explain the DAMA effect, so we assume
$\tilde{\eta}^{est}(v_{min}>v_2^{DAMA,Na})$=0 (this is also consistent
to dominance of scatterings on sodium in DAMA in the mass range
$m_{WIMP}<60$ GeV, as explained above); iv) we then use
$\tilde{\eta}^{est}(v_{min})$ to directly calculate for each
experiment among $k$=SIMPLE, COUPP, PICASSO and PICO-2L and for each
energy threshold $E_{th,i}$ the expected number of WIMP events
$N_{k,i}^{expected}$ and compare it to the corresponding upper bound
$N_{k,i}^{bound}$ (see Appendix \ref{app:exp} for further
details). Then, a straightforward generalization of the compatibility
factor of Eq.(\ref{eq:compatibility_factor_eta_i}) is:

\be {\cal D}(m_{WIMP},r_i)\rightarrow \max \left ({\cal
  D}(m_{WIMP},r_i), \frac{N_{k,i}^{expected}}{N_{k,i}^{bound}} \right ).
\label{eq:compatibility_factor_generilized}
\ee

Notice that the above procedure is very general, since it allows to
get a halo--independent bound in the case of any experiment that does
not observe a signal. In particular this means that the requirement
that one target dominates the expected rate is only needed to get
estimations of the halo function from those experiments that observe
an excess, but is not indispensable to get a
conservative bound for those reporting a null result. In the following Section we will use the above
definition of the compatibility factor to explore the WIMP
parameter space for the generalized spin--dependent interactions introduced in Section \ref{sec:eft}.

\section{Results}
\label{sec:results}

Each of the seven phenomenological models ${\cal O}_i$ (with $i$=4, 6,
7, 9, 10, 14, 46) listed at the end of Section \ref{sec:eft} depends on three
parameters: the WIMP mass $m_{WIMP}$ and the two components
$(c_i^p,c_i^n)$ of the corresponding effective coupling. However,
following the halo--independent approach summarized in Section
\ref{sec:halo_independent}, we get our estimations of the
$\tilde{\eta}$ or $\tilde{\xi}$ functions factorizing out the
reference cross section $\sigma_{ref}=c^p_i\mu_{\chi N}/\pi$. As a
consequence of this the coupling $c^p_i$ cancels out in our definition
(\ref{eq:compatibility_factor_eta_i},\ref{eq:compatibility_factor_generilized})
of the compatibility factor ${\cal D}$. So in each model the
phenomenology depends only on two parameters: $m_{WIMP}$ and the ratio
$r_i$ (in the case of model ${\cal O}_{46}$ we
factorize $\sigma_{ref}=c^p_6\mu_{\chi N}/\pi$).

Moreover, the pairs of models ${\cal O}_4$--${\cal O}_7$ and ${\cal
  O}_9$--${\cal O}_{14}$ only differ for an additional
$(v_T^{\perp})^2$ factor in the cross section (see Table
\ref{table:ref_nref}), which is absorbed in the definition of the halo
function $\tilde{\xi}$. So in each pair the
two models have identical compatibility factors ${\cal D}$ (in each
case the estimations $\tilde{\xi}_i$ for the velocity--dependent model
are simply rescaled by a common factor with respect to the
corresponding $\tilde{\eta}_i$ in the velocity--independent one) . The
bottom line is that only five relevant cases remain: ${\cal O}_4$,
${\cal O}_6$, ${\cal O}_9$, ${\cal O}_{10}$ and ${\cal O}_{46}$.

\begin{figure}
\begin{center}
\includegraphics[width=0.49\columnwidth, bb=73 193 513 636]{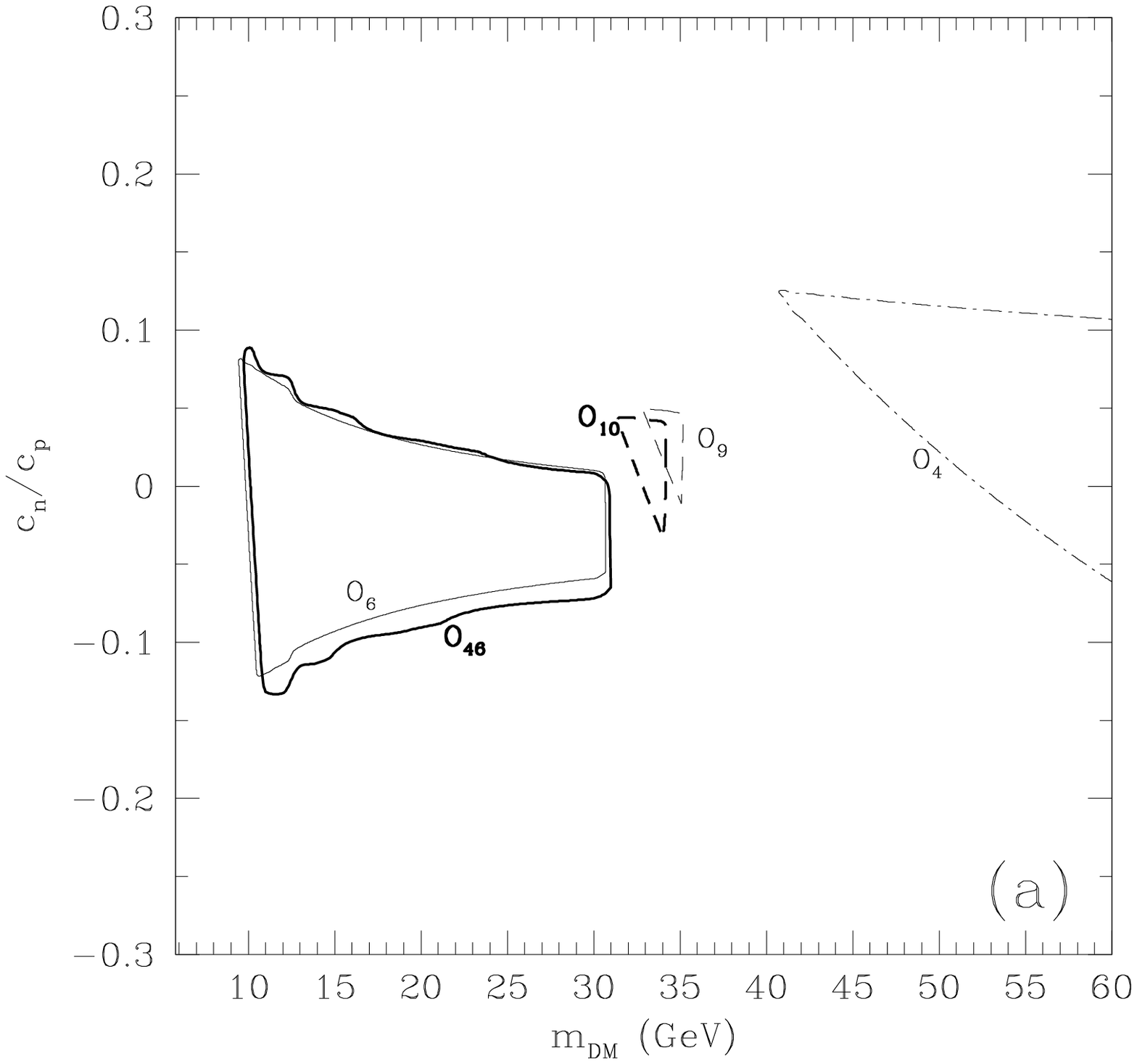}
\includegraphics[width=0.49\columnwidth, bb=73 193 513 636]{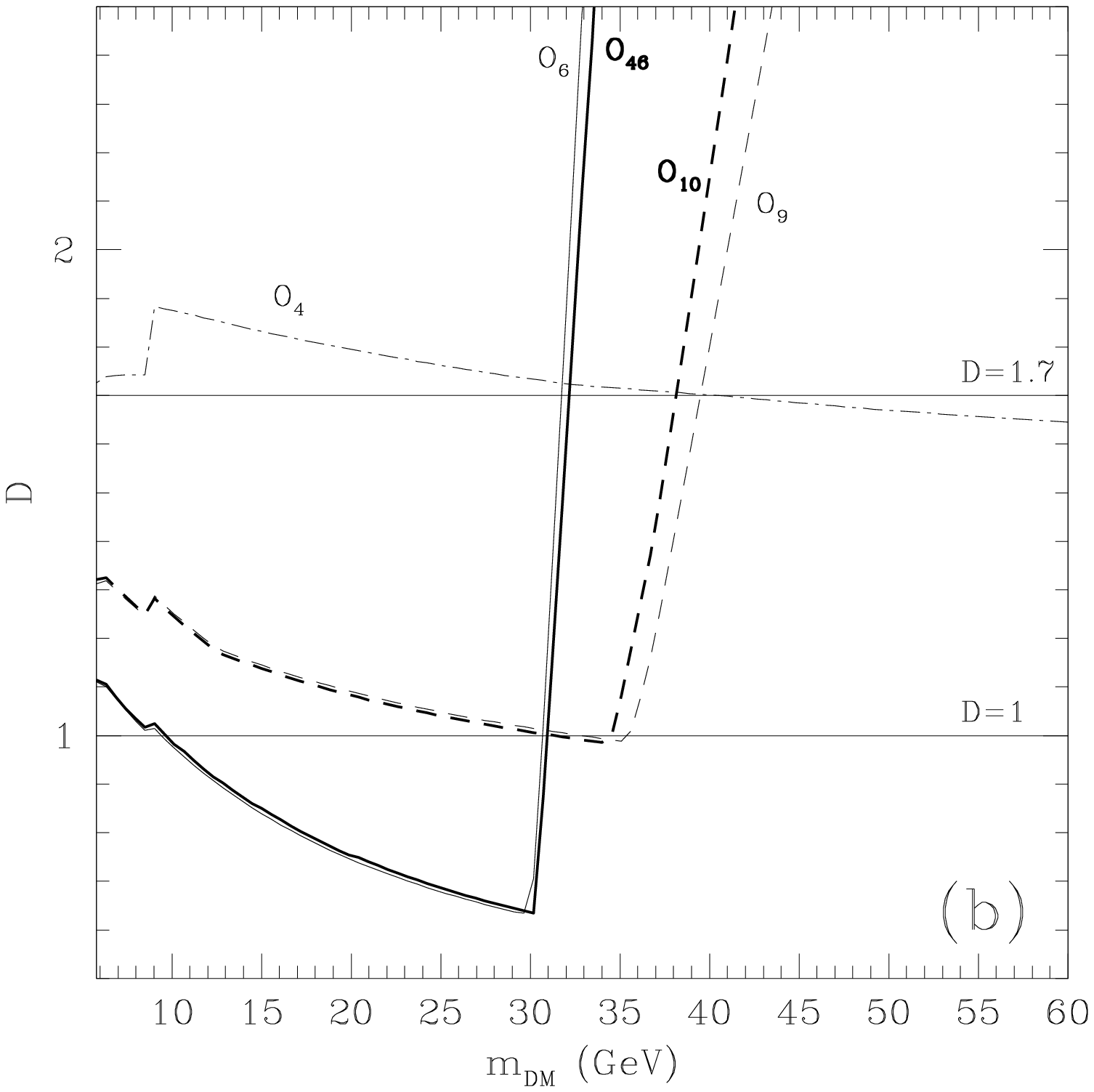}
\end{center}
\caption{{\bf (a)} Contour plot in the $m_{WIMP}$--$c^n_i/c^p_i$ plane
  for the compatibility factor ${\cal D}$ defined in
  Eqs. (\ref{eq:compatibility_factor_eta_i},\ref{eq:compatibility_factor_generilized}).
  The constant value ${\cal D}$=1 is shown for models ${\cal O}_i$,
  $i=6,46,9,10$, while ${\cal D}$=1.7 is plotted for ${\cal
    O}_4$. {\bf (b)} For the same models ${\cal O}_i$ the minimum of
  ${\cal D}$ is plotted as a function of $m_{WIMP}$ when $c^n_i/c^p_i$
  is marginalized. }
\label{fig:d}
\end{figure}

The results of our analysis is shown in Fig.\ref{fig:d}. The
left--hand panel shows the contour plot in the $m_{WIMP}$--$r_i$ plane
for ${\cal D}$=1 of models ${\cal O}_i$, $i=6,46,9,10$ and ${\cal
  D}$=1.7 for ${\cal O}_4$; the right--hand panel shows the minimum of
${\cal D}$ as a function of $m_{WIMP}$ when $r_i$ is marginalized. A
few features arise from both plots: i) when $c^n/c^p\rightarrow$0 some
intervals of $m_{WIMP}$ exist where the DAMA effect is compatible to
constraints from other direct--detection experiments for generalized
spin--dependent interactions (${\cal O}_i$, $i=6,46,9,10$), while the
compatibility factor for the standard spin--dependent interaction
${\cal O}_4$ is always above ${\cal D}$=1; ii) from Table
\ref{table:ref_nref} one can see that the pairs of models ${\cal
  O}_6$--${\cal O}_{46}$ and ${\cal O}_9$--${\cal O}_{10}$ have the
same momentum dependence, while they differ for the nuclear response
function: since however $W_{\Sigma^{\prime}}(q^2)\simeq 2
W_{\Sigma^{\prime\prime}}(q^2)$ (at least for small $q^2$) the
compatibility factor ${\cal D}$ has very similar behaviours for the
models in each pair ; ii) the tension between DAMA and other
experiments is better alleviated in those models where momentum
dependence is largest: in fact, models ${\cal O}_6$ and ${\cal
  O}_{46}$, which reach the best compatibility, depend on momentum
through a factor $q^4$ compared to ${\cal O}_9$ and ${\cal O}_{10}$
where the dependence is through $q^2$ (see again Table
\ref{table:ref_nref}); iv) for $m_{WIMP}\gsim$ 30 GeV the
compatibility factor rises steeply for models ${\cal O}_6$, ${\cal
  O}_{46}$, ${\cal O}_9$ and ${\cal O}_{10}$, while no such feature is
observed for the case ${\cal O}_4$.

The last two properties can be understood in the following way.  If
$m_{WIMP}\lsim$ 30 GeV, the WIMP signal in bubble chambers and droplet
detectors is dominated by scatterings off fluorine. In particular, in
this $m_{WIMP}$ range in order to deposit recoil energies above
threshold scatterings off iodine in COUPP require $v_{min}$ values
beyond the corresponding range for the DAMA signal, and we make the
conservative assumption that in this case the halo function
$\tilde{\eta}$ vanishes. Moreover scatterings off chlorine in SIMPLE
are subdominant due to the suppressed nuclear response function. In
this case the transferred momenta $q^2$ which explain the DAMA
modulation effect for WIMP scatterings off sodium in DAMA are larger
than the corresponding ones off fluorine in COUPP, PICASSO and
PICO-2L. For instance, for $m_{WIMP}$=25 GeV one has 285 MeV$^2\lsim
(q_{DAMA}^{Na})^2\lsim $570 MeV$^2$, 275 MeV$^2 \lsim
(q_{COUPP}^{F})^2 \lsim $470 MeV$^2$, 60 MeV$^2 \lsim
(q_{PICASSO}^{F})^2 \lsim$470 MeV$^2$, 113 MeV$^2\lsim
(q_{PICO-2L}^{F})^2\lsim$470 MeV$^2$.  This implies that models where
the expected detection rate depends on one additional factor
$(q^2)^n$, $n>0$ such as ${\cal O}_i$, $i=6,46,9,10$ present a
relative enhancement of the expected rate in DAMA compared to that in
fluorine detectors, with a consequent relative loss of sensitivity for
the latter.  An exception to this argument is SIMPLE, where the $q^2$
interval for scatterings off fluorine has more overlap with that of
DAMA, 283 MeV$^2<q_{SIMPLE}^{F}<$470 MeV$^2$, but which is overall
less constraining than the other detectors due to the lower exposure
(see Appendix \ref{app:exp}).  On the other hand, for $m_{WIMP}\gsim$
30 GeV scatterings off iodine in COUPP become kinematically allowed,
with values of the transferred momenta which are much larger than
those related to fluorine and sodium. For instance, for $m_{WIMP}$=35
GeV one has 1850 MeV$^2\lsim (q_{COUPP}^{I})^2\lsim $2350
MeV$^2$. Clearly, this implies a strong enhancement of the expected
signal in COUPP for interactions involving an additional dependence on
$q^2$, with a consequent steep rise of the compatibility factor ${\cal
  D}$, as observed in Fig. \ref{fig:d}.  This effect is not present
for the standard spin--dependent interaction ${\cal O}_4$, whose
expected rate has no $q^2$ dependence.

\begin{figure}
\begin{center}
\includegraphics[width=0.49\columnwidth, bb=73 193 513 636]{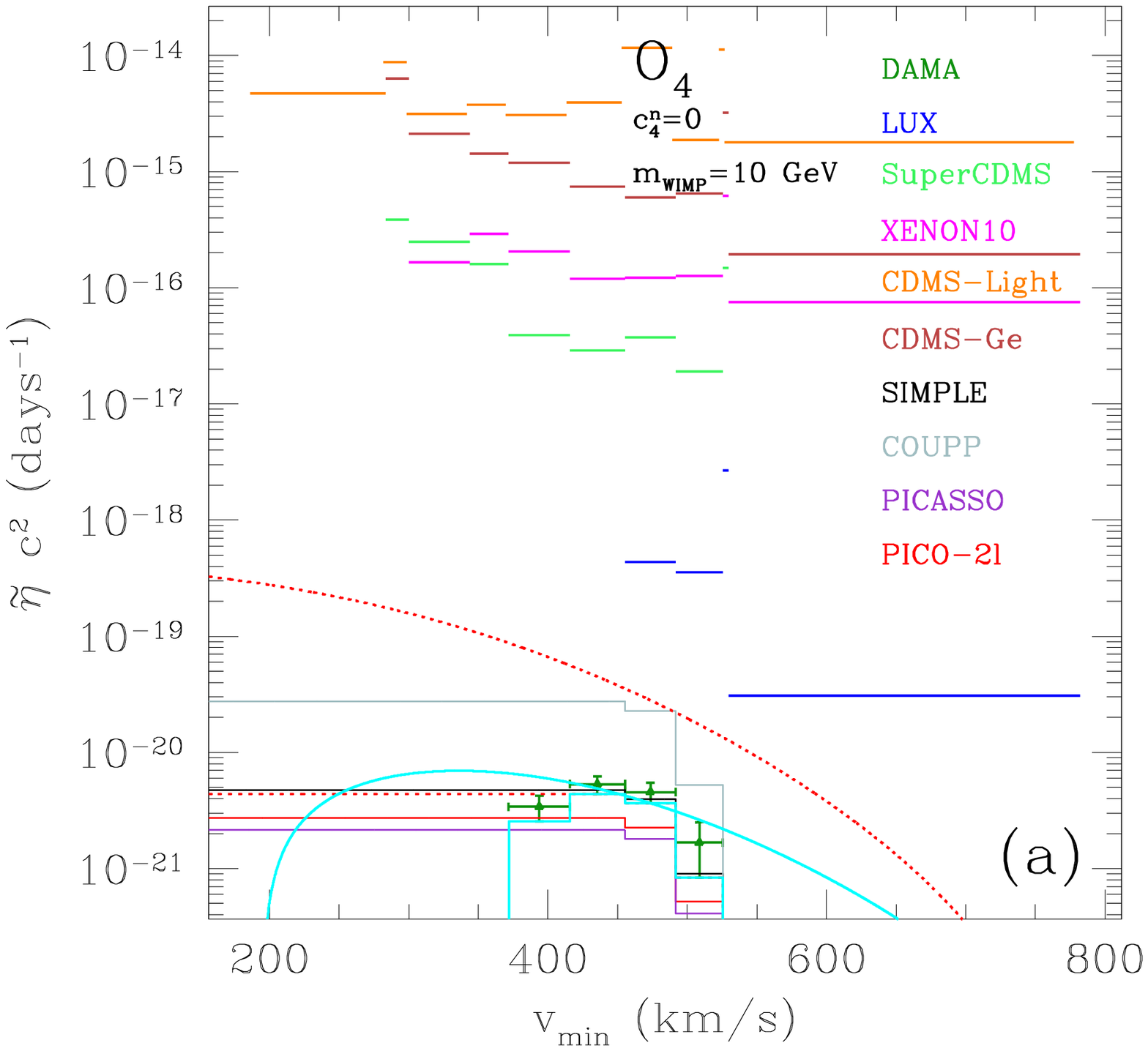}
\includegraphics[width=0.49\columnwidth, bb=73 193 513 636]{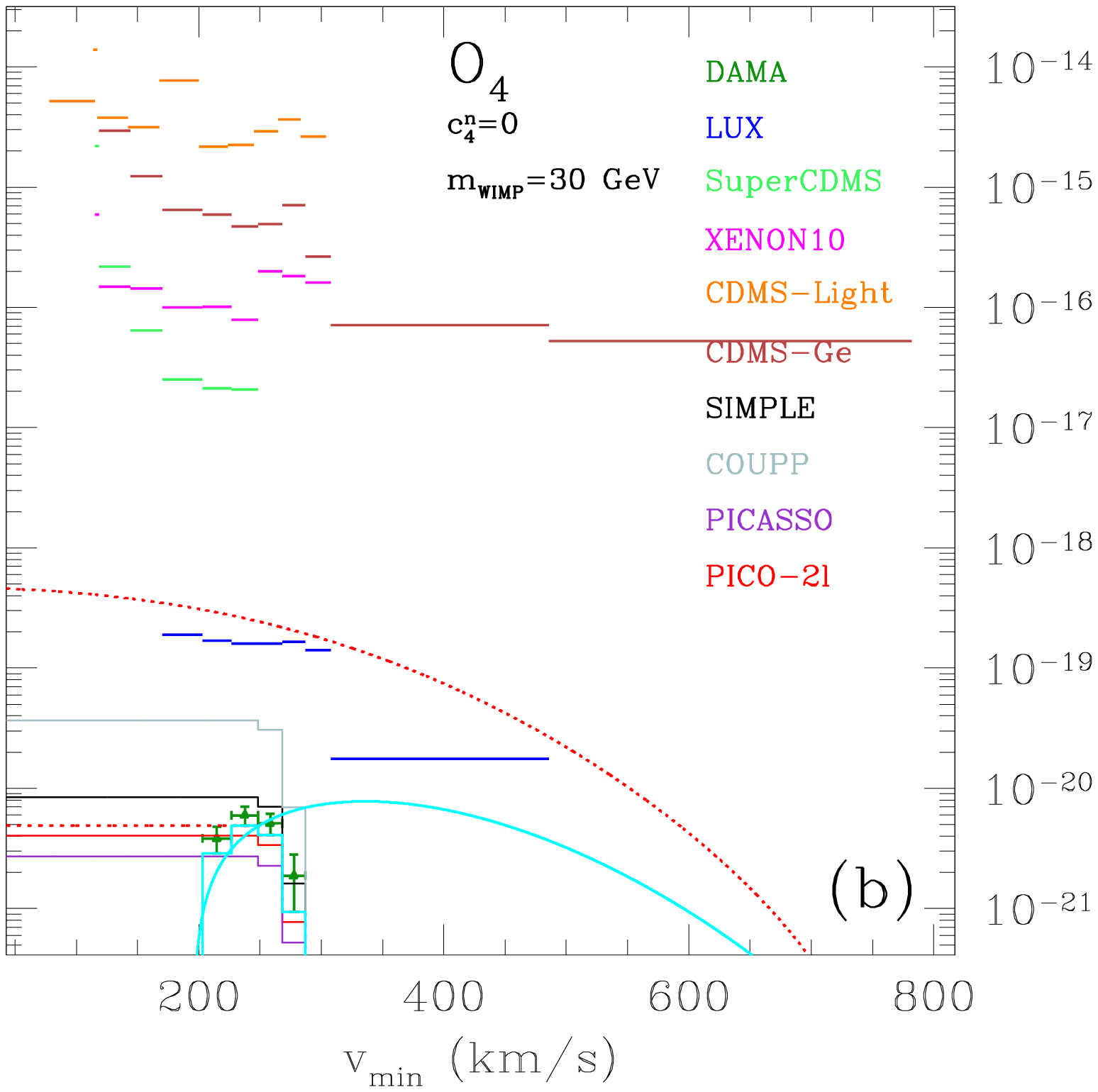}
\end{center}
\caption{Measurements and bounds for the function $\tilde{\eta}$
  defined in Eq. (\ref{eq:eta}) for model ${\cal O}_4$ (standard
  spin--dependent interaction) and $c^n_4$=0. {\bf (a)} $m_{WIMP}$=10
  GeV; {\bf (b)} $m_{WIMP}$=30 GeV. The $\tilde{\xi}$ determinations
  for model ${\cal O}_7$ would be rescaled by an approximately common
  factor $\simeq$ 3/2 with respect to the $\tilde{\eta}$ values shown
  in this figure.}
\label{fig:eta_vmin_c4}
\end{figure}

\begin{figure}
\begin{center}
\includegraphics[width=0.49\columnwidth, bb=73 193 513 636]{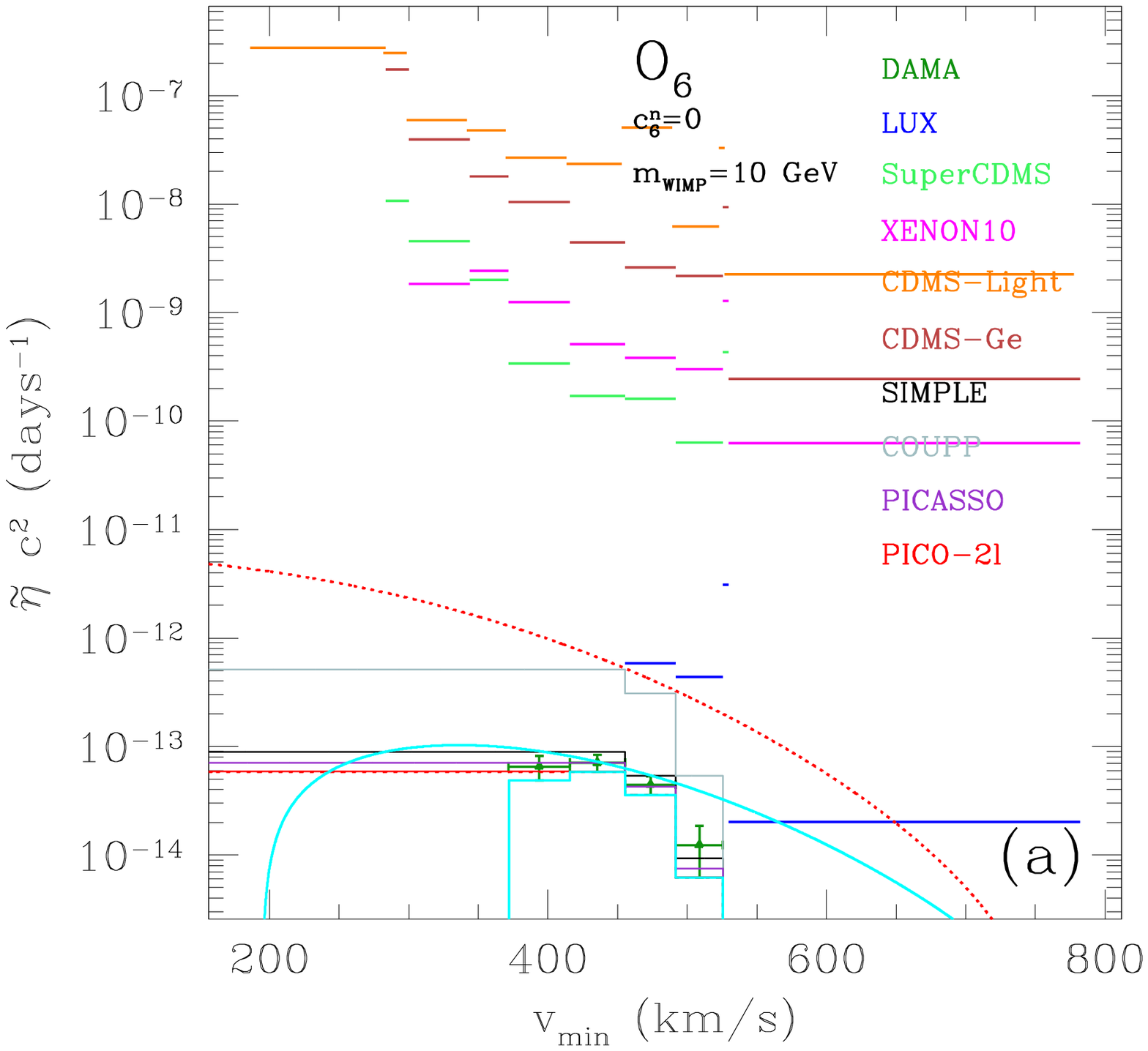}
\includegraphics[width=0.49\columnwidth, bb=73 193 513 636]{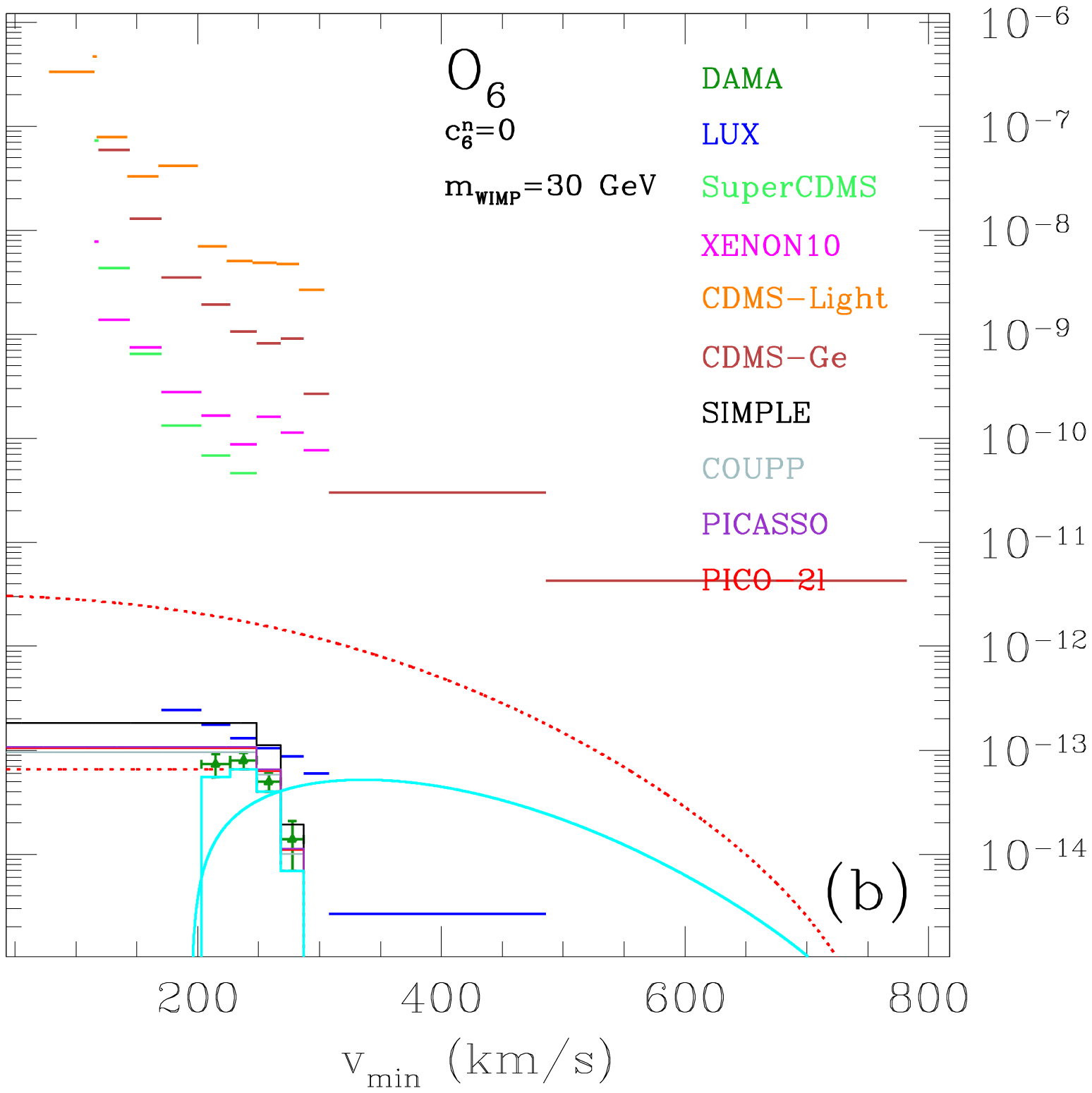}
\end{center}
\caption{The same as in Figure \ref{fig:eta_vmin_c4} for model ${\cal
    O}_6$. The $\tilde{\eta}$ determinations
  for model ${\cal O}_{46}$ would be rescaled by an approximately common
  factor $\simeq$ 2 with respect to the $\tilde{\eta}$ values shown
  in this figure.}
\label{fig:eta_vmin_c6}
\end{figure}

\begin{figure}
\begin{center}
\includegraphics[width=0.49\columnwidth, bb=73 193 513 636]{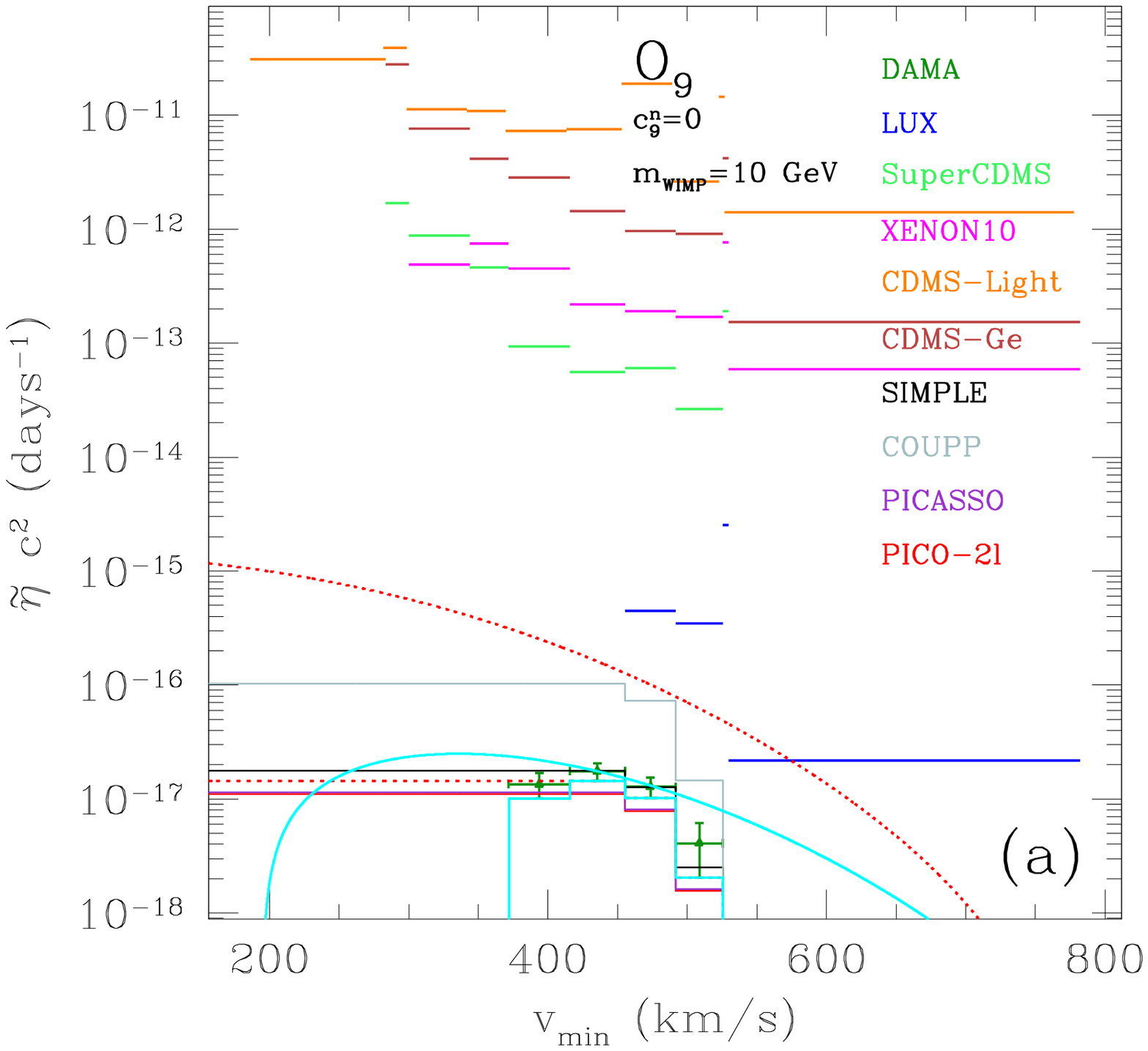}
\includegraphics[width=0.49\columnwidth, bb=73 193 513 636]{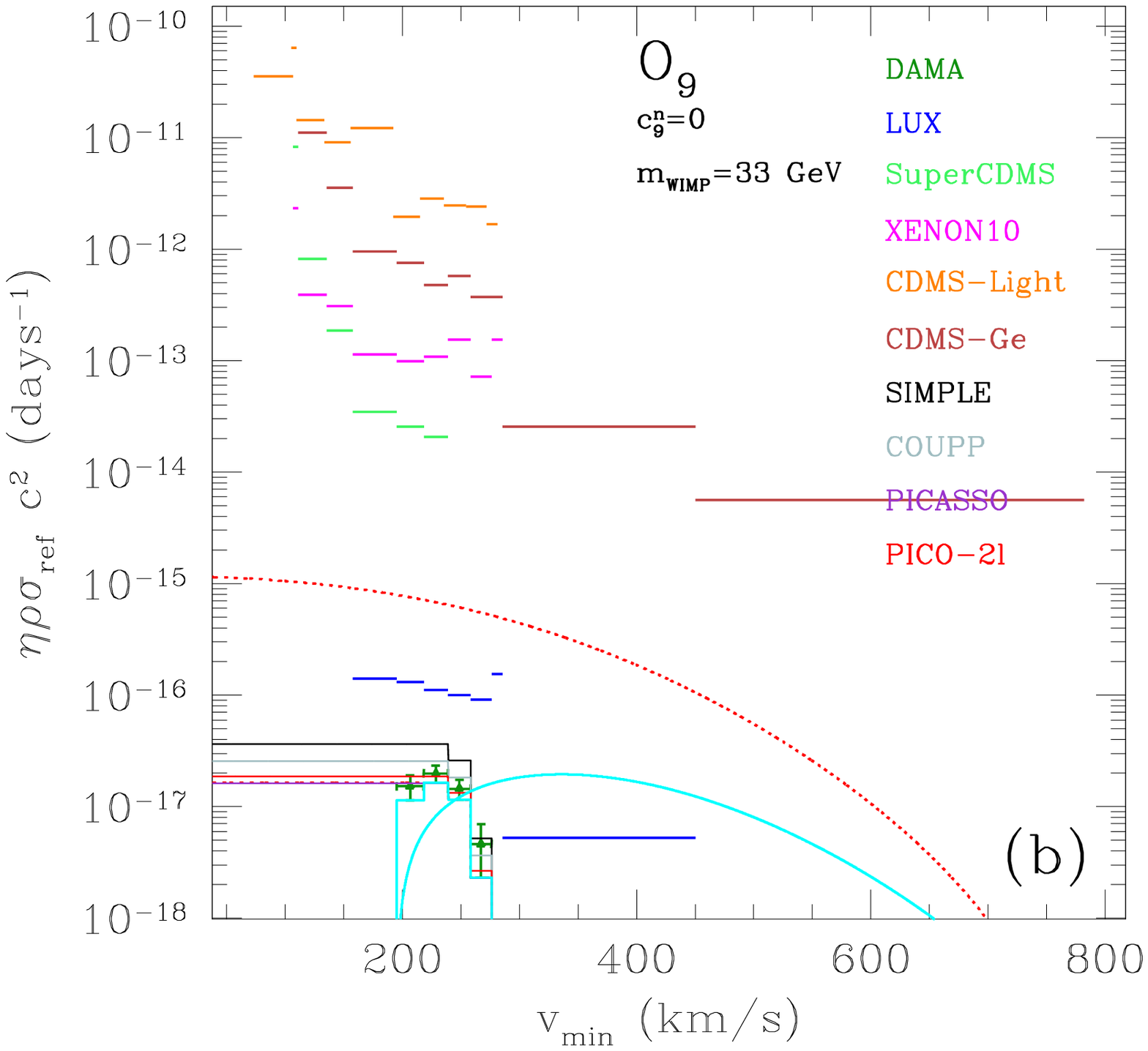}
\end{center}
\caption{ The same as in Figure \ref{fig:eta_vmin_c4} for model ${\cal
    O}_9$. The $\tilde{\xi}$ determinations for model ${\cal O}_{14}$
  would be rescaled by a common factor 2 with respect to the
  $\tilde{\eta}$ values shown in this figure. The $\tilde{\eta}$
  determinations for model ${\cal O}_{10}$ would be rescaled by an
  approximately common factor $\simeq$ 1/2.}
\label{fig:eta_vmin_c9}
\end{figure}

In order to discuss in more detail the phenomenology in Figures
\ref{fig:eta_vmin_c4}, \ref{fig:eta_vmin_c6} and \ref{fig:eta_vmin_c9}
the measurements and bounds for the function $\tilde{\eta}$ obtained
using Eq. (\ref{eq:eta_bar_vt}) are plotted as a function of
$v_{min}$. In particular, in all these figures we take
$r_i=0$.  Fig. \ref{fig:eta_vmin_c4}(a) shows the case of a
standard spin--dependent interaction ${\cal O}_4$ when $m_{WIMP}$=10
GeV, while Fig. \ref{fig:eta_vmin_c4}(b) shows the same when
$m_{WIMP}$=30 GeV. On the other hand, Figure \ref{fig:eta_vmin_c6}(a)
shows the case of model ${\cal O}_6$ for $m_{WIMP}$=10 GeV while
Fig. \ref{fig:eta_vmin_c6}(b) shows the same when $m_{WIMP}$=30 GeV.
Finally, in Figure \ref{fig:eta_vmin_c4}(a) and
\ref{fig:eta_vmin_c4}(b) model ${\cal O}_9$ is shown for
$m_{WIMP}$=10 GeV and $m_{WIMP}$=33 GeV, respectively.

As discussed before, phenomenology of these models is practically
degenerate with other scenarios. In particular, the $\tilde{\xi}$
determinations for model ${\cal O}_7$ would be rescaled by an
approximately common factor ($\simeq$ 3/2) with respect to the
$\tilde{\eta}$ shown in figure \ref{fig:eta_vmin_c4} for ${\cal O}_4$.
In the same way the phenomenology of ${\cal O}_{46}$ is very similar
to model ${\cal O}_{6}$, apart from an overall approximate factor
($\simeq$ 1/2) in the $\tilde{\eta}$ determinations, so it can be
described by Figure \ref{fig:eta_vmin_c6}.  Moreover, the
$\tilde{\xi}$ determinations for model ${\cal O}_{14}$ would be
rescaled by an exact common factor of 2 with respect to the
$\tilde{\eta}$ values shown in Figure \ref{fig:eta_vmin_c9} for ${\cal
  O}_{9}$; on the other hand, the $\tilde{\eta}$ determinations for
model ${\cal O}_{10}$ would be rescaled by an approximately common
factor $\simeq$ 1/2.

From all these plots we observe that germanium detectors and LUX imply
relatively lax constraints, due to the suppression of their nuclear
response functions when $c^n_i$=0 (and, in the case of germanium, for
the small natural abundance of $^{73}Ge$, the only isotope carrying
spin).

Moreover, in each plot the step--like cyan curve represents our
conservative estimation $\tilde{\eta}^{est}_1(v_{min})$ of the minimal
$\tilde{\eta}_1$ modulated halo function compatible to the DAMA
signal. In particular, as explained before, we assume that
$\tilde{\eta}^{est}_1(v_{min})$ vanishes outside the DAMA $v_{min}$
interval.  The corresponding estimation $\tilde{\eta}^{est}(v_{min})\ge \tilde{\eta}^{est}_1(v_{min})$
of the minimal $\tilde{\eta}$ function is obtained by requiring it to be a decreasing
function of $v_{min}$ and is represented by the piecewise dotted red
line.

As explained before, we use the function $\tilde{\eta}^{est}(v_{min})$
to calculate the expected rate on droplet detectors and bubble
chambers in the compatibility factor of
Eq.(\ref{eq:compatibility_factor_generilized}). We decide to show the
corresponding constraints by plotting for each experiment the function
$\tilde{\eta}^{est}(v_{min})$ rescaled by a constant factor in such a
way that the corresponding expected rate is equal to the corresponding
most constraining upper bound on the number of events. As can be seen
from all the figures, also in the most favorable situations (${\cal
  O}_6\simeq {\cal O}_{46}$ for $m_{WIMP}\simeq$ 30 GeV) the bound is
always relatively close (within one sigma) to the minimal DAMA signal.
Since the DAMA estimations are for $\tilde{\eta}_1$ while the bounds
are for $\tilde{\eta}$ this implies that compatibility among them
necessarily requires a modulation amplitude fraction in DAMA much
larger then that predicted for a standard Isothermal Sphere (which is
typically below 10\%).  In the same figures the continuous cyan line
and dotted red line represent $\tilde{\eta}_1^{Maxwellian}(v_{min})$
and $\tilde{\eta}^{Maxwellian}(v_{min})$ for the case of a Maxwellian
velocity distribution, where $\tilde{\eta}_1^{Maxwellian}(v_{min})$ is
normalized in such a way to minimize a $\chi$--square with the DAMA
modulation amplitudes. Indeed, the red dotted line is well above
existing constraints on $\tilde{\eta}$ in all cases. Moreover, notice
that a Maxwellian for $m_{WIMP}$=30 GeV predicts a sizeable
contribution to the event rate due to scatterings off iodine, so this
explain the particularly poor fit with the $\tilde{\eta}$ estimations
for DAMA, where dominance on sodium was assumed.

\section{Conclusions}
\label{sec:conclusions}

In the present paper we have used non--relativistic Effective Field
Theory to classify the most general spin--dependent WIMP--nucleus
interactions, and within this class of models we have discussed the
viability of an interpretation of the DAMA modulation result in terms
of a WIMP signal, using a halo--independent approach in which all
dependencies from astrophysics are factorized in a single halo
function.

One of the main motivations of the spin--dependent scenario is the
fact that the most stringent bounds on the interpretation of the DAMA
effect in terms of WIMP--nuclei scatterings arise today from detectors
using xenon (XENON100, LUX) and germanium (CDMS) whose spin is mostly
originated by an unpaired neutron, while both sodium and iodine in
DAMA have an unpaired proton: if the WIMP effective coupling to
neutrons is suppressed compared to that on protons this class of
bounds can be evaded. In this case the most constraining remaining
bounds arise from droplet detectors (SIMPLE, COUPP) and bubble
chambers (PICASSO, PICO-2L) , which all use nuclear targets (fluorine,
chlorine and iodine) with an unpaired proton.

From the phenomenological point of view we found that, although
several relativistic Effective Field Theories can lead to a
spin--dependent cross section, in some cases involving an explicit
dependence of the scattering cross section on the WIMP incoming
velocity (see Table \ref{table:ref_nref}), three main scenarios can be
singled out in the non--relativistic limit which (approximately)
encompass them all, and that only differ by their explicit dependence
on the transferred momentum, $(q^2)^n$, $n$=0,1,2: using the notation
of Eq.(\ref{eq:ops}), they are ${\cal O}_4$, ${\cal O}_9$ and ${\cal
  O}_6$, respectively.

In our quantitative analysis we pointed out that the requirement for a
halo--independent analysis that one target dominates the expected rate
is only needed to get estimations of the halo function from those
experiments that observe an excess, but is not indispensable to get a
conservative bound for those reporting a null result. We achieved this
by adopting the procedure to find the minimal halo function
$\tilde{\eta}^{est}(v_{min})$ compatible to the DAMA signal and then
use it to calculate expected rates in other experiments, including
droplet detectors and bubble chambers which contain several target
nuclei. Using this approach we also concluded that scatterings on
iodine can be assumed to be below threshold in DAMA as long as
$m_{WIMP}\lsim$ 60 GeV, allowing the factorization of the halo
function for sodium and evading the KIMS bound (also using iodine) in
the same WIMP mass range.

In particular, we found that, for $m_{WIMP}\lsim 30$ GeV and with our
assumptions on $\tilde{\eta}^{est}(v_{min})$, the WIMP signal in
bubble chambers and droplet detectors is dominated by scatterings off
fluorine. In this case models where the expected detection rate
depends on one additional factor $(q^2)^n$, $n>0$ show a relative
enhancement of the expected rate in DAMA compared to that in fluorine
detectors, with a consequent relative loss of sensitivity for the
latter, because the transferred momenta $q^2$ which explain the DAMA
modulation effect for WIMP scatterings off sodium in DAMA are larger
than the corresponding ones off fluorine.  In this way compatibility
between DAMA and other constraints can be achieved for ${\cal O}_6$
and, to a lesser extent, for ${\cal O}_9$, but not for the standard
spin--dependent scenario ${\cal O}_4$.  These conclusions are only
valid for a WIMP velocity distribution in the halo of our Galaxy which
departs from a Maxwellian.  On the other hand, for $m_{WIMP}\gsim 30$ GeV
a strong tension between DAMA and COUPP arises for both ${\cal O}_6$
and ${\cal O}_9$ because scatterings off iodine in COUPP become
kinematically allowed, with values of the transferred momenta which
are much larger than those related to fluorine and sodium.

\acknowledgments
This work was supported by the National Research
Foundation of Korea(NRF) grant funded by the Korea government(MOE)
(No. 2011-0024836).

\appendix
\section{WIMP response functions}
\label{app:wimp_eft}

We collect here the WIMP particle--physics response functions introduced in Eq.(\ref{eq:squared_amplitude}) and adapted from \cite{haxton}:
\begin{eqnarray}
 R_{M}^{\tau \tau^\prime}\left(v_T^{\perp 2}, {q^2 \over m_N^2}\right) &=& c_1^\tau c_1^{\tau^\prime } + {j_\chi (j_\chi+1) \over 3} \left[ {q^2 \over m_N^2} v_T^{\perp 2} c_5^\tau c_5^{\tau^\prime }+v_T^{\perp 2}c_8^\tau c_8^{\tau^\prime }
+ {q^2 \over m_N^2} c_{11}^\tau c_{11}^{\tau^\prime } \right] \nonumber \\
 R_{\Phi^{\prime \prime}}^{\tau \tau^\prime}\left(v_T^{\perp 2}, {q^2 \over m_N^2}\right) &=& \left [{q^2 \over 4 m_N^2} c_3^\tau c_3^{\tau^\prime } + {j_\chi (j_\chi+1) \over 12} \left( c_{12}^\tau-{q^2 \over m_N^2} c_{15}^\tau\right) \left( c_{12}^{\tau^\prime }-{q^2 \over m_N^2}c_{15}^{\tau^\prime} \right)\right ]\frac{q^2}{m_N^2}  \nonumber \\
 R_{\Phi^{\prime \prime} M}^{\tau \tau^\prime}\left(v_T^{\perp 2}, {q^2 \over m_N^2}\right) &=& \left [ c_3^\tau c_1^{\tau^\prime } + {j_\chi (j_\chi+1) \over 3} \left( c_{12}^\tau -{q^2 \over m_N^2} c_{15}^\tau \right) c_{11}^{\tau^\prime }\right ] \frac{q^2}{m_N^2} \nonumber \\
  R_{\tilde{\Phi}^\prime}^{\tau \tau^\prime}\left(v_T^{\perp 2}, {q^2 \over m_N^2}\right) &=&\left [{j_\chi (j_\chi+1) \over 12} \left ( c_{12}^\tau c_{12}^{\tau^\prime }+{q^2 \over m_N^2}  c_{13}^\tau c_{13}^{\tau^\prime}  \right )\right ]\frac{q^2}{m_N^2} \nonumber \\
   R_{\Sigma^{\prime \prime}}^{\tau \tau^\prime}\left(v_T^{\perp 2}, {q^2 \over m_N^2}\right)  &=&{q^2 \over 4 m_N^2} c_{10}^\tau  c_{10}^{\tau^\prime } +
  {j_\chi (j_\chi+1) \over 12} \left[ c_4^\tau c_4^{\tau^\prime} + \right.  \nonumber \\
 && \left. {q^2 \over m_N^2} ( c_4^\tau c_6^{\tau^\prime }+c_6^\tau c_4^{\tau^\prime })+
 {q^4 \over m_N^4} c_{6}^\tau c_{6}^{\tau^\prime } +v_T^{\perp 2} c_{12}^\tau c_{12}^{\tau^\prime }+{q^2 \over m_N^2} v_T^{\perp 2} c_{13}^\tau c_{13}^{\tau^\prime } \right] \nonumber \\
    R_{\Sigma^\prime}^{\tau \tau^\prime}\left(v_T^{\perp 2}, {q^2 \over m_N^2}\right)  &=&{1 \over 8} \left[ {q^2 \over  m_N^2}  v_T^{\perp 2} c_{3}^\tau  c_{3}^{\tau^\prime } + v_T^{\perp 2}  c_{7}^\tau  c_{7}^{\tau^\prime }  \right]
       + {j_\chi (j_\chi+1) \over 12} \left[ c_4^\tau c_4^{\tau^\prime} +  \right.\nonumber \\
       &&\left. {q^2 \over m_N^2} c_9^\tau c_9^{\tau^\prime }+{v_T^{\perp 2} \over 2} \left(c_{12}^\tau-{q^2 \over m_N^2}c_{15}^\tau \right) \left( c_{12}^{\tau^\prime }-{q^2 \over m_N^2}c_{15}^{\tau \prime} \right) +{q^2 \over 2 m_N^2} v_T^{\perp 2}  c_{14}^\tau c_{14}^{\tau^\prime } \right] \nonumber \\
     R_{\Delta}^{\tau \tau^\prime}\left(v_T^{\perp 2}, {q^2 \over m_N^2}\right)&=& {j_\chi (j_\chi+1) \over 3} \left( {q^2 \over m_N^2} c_{5}^\tau c_{5}^{\tau^\prime }+ c_{8}^\tau c_{8}^{\tau^\prime } \right)\frac{q^2}{m_N^2} \nonumber \\
 R_{\Delta \Sigma^\prime}^{\tau \tau^\prime}\left(v_T^{\perp 2}, {q^2 \over m_N^2}\right)&=& {j_\chi (j_\chi+1) \over 3} \left (c_{5}^\tau c_{4}^{\tau^\prime }-c_8^\tau c_9^{\tau^\prime} \right) \frac{q^2}{m_N^2}.
\label{eq:wimp_response_functions}
\end{eqnarray}

\section{Experimental inputs for the analysis}
\label{app:exp}

In this Appendix we summarize the experimental inputs that we have
used to evaluate the response function defined in Eq. (\ref{eq:core})
for each of the experiments included in our analysis. Whenever
applicable we will follow the convention to indicate with $E_R$ the
true recoil energy, with $E_{ee}$ the electron--equivalent energy
($E_{ee}=Q(E_R) E_R$ with $Q$ the quenching factor) and with
$E^{\prime}$ the visible energy, as introduced in Section
\ref{sec:halo_independent}. In the case of bolometric measurements we
assume $Q=1$.  With the exceptions of LUX we model the energy
resolution with a Gaussian and we indicate the corresponding variance.

{\bf DAMA} We have taken the modulation amplitudes in 0.5 keVee bins
from Fig.6 of Ref.\cite{dama} (already normalized to counts/day/kg/keV
for a total exposure of 1.17 ton yr), adopting the signal region 2
keVee $\le E^{\prime}\le$ 4 keVee. We have adopted the value
$Q_{Na}$=0.3 for the quenching factor of sodium.

{\bf LUX} In the case of LUX we have assumed zero WIMP candidate
events in the range 2 PE$\le S_1\le$30 PE in the lower half of the
signal band, as shown in Fig. 4 of Ref. \cite{lux} for the primary
scintillation signal $S_1$ (directly in Photo Electrons, PE) for an
exposure of 85.3 days and a fiducial volume of 118 kg of xenon.
Following Ref. \cite{xenon100_response} (see Eqs. (14-15)) we have
modeled the detector's response with a Poissonian fluctuation of the
$S_1$ scintillation photoelectrons combined with a Gaussian resolution
$\sigma_{PMT}$=0.5 PE for the photomultiplier so that the response
functions defined in Eq.(\ref{eq:core}) are modified into:

\begin{eqnarray}
&&{\cal R}_{[\Ed_1,\Ed_2]}^m=
\sum_T 8 N_T \mu_{T}^2\frac{1}{\sigma_{ref}}\sum_{k}\sum_{\tau\tau^{\prime}}R^{\tau\tau^{\prime}}_{m k}\left[q^2(v_{min}) \right]
\frac{W^{\tau\tau^{\prime}}_{T k} \left [y(v_{min}) \right ]}{2 j_T+1}\times \nonumber\\
&&\int_{S_{1,min}}^{S_{1,max}}dS_1 \, \sum_{n=1}^{\infty} Gauss\left (S_1 | n,\sqrt{n} \sigma_{PMT} \right )
Poiss\left [n,\nu(E_R) \right ]\xi_{cuts}(S_1)
,\;\;\;\;m=0,1.
\label{eq:core_liquid}
\end{eqnarray}

\noindent In the equation above
$Poiss(n,\lambda)=\lambda^n/n!\exp(-\lambda)$, while $\xi_{cuts}$
represents the combination of a 50\% acceptance combined with the
quality cut efficiency (taken from Fig. 9 of Ref. \cite{lux}). Moreover the
expected number of PE for a given recoil energy $E_R$ is given by:

\begin{equation}
  \nu(E_R)=E_{R}\times L_{eff}(E_R) \times L_y,
\label{eq:nu}
\end{equation}

\noindent with $L_y$=8.8 PE. We have taken $L_{eff}(E_R)$ from
\cite{lux_slides} where it is calculated including the effect of the
electric field, and assumed to vanish for
$S_1<$ 3 PE.

{\bf SuperCDMS} We include the low--energy analysis of
SuperCDMS\cite{super_cdms} with a germanium target in the energy
range 1.6 keVnr$<E_R<$ 10 keVnr with a total exposition of 577 kg day
and 11 observed WIMP candidates. The energy resolution is given by
$\sigma_{CDMS-Si}(E^{\prime})=\sqrt{0.293^2+0.056^2
  (E^{\prime}/\mbox{keVnr})}$ in keVnr\cite{cdms_resolution}

{\bf XENON10} The analysis of XENON10 makes use of the secondary
ionization signal $S_2$ only, with an exposition of 12.5 day and a
fiducial mass of 1.2 kg. We take the scale of the recoil energy $E_R$
and the recorded event spectrum in the energy range 1.4 keVnr$<E_R<$
10 keVnr directly from Fig. 2 of Ref. \cite{xenon10}. The energy
resolution is given by: $\sigma_{XENON10}=E_R/\sqrt{E_R Q_y(E_R)}$
where $Q_y(E_R)$ is the electron yield that we calculate with the same
choice of parameters as in Fig. 1 of \cite{xenon10}.

{\bf CDMSlite} CDMSlite\cite{cdms_lite} analyzes the very low range
0.170 keVee$<E_{ee}<$7 keVee for the electron--equivalent energy using a
fiducial mass of 0.6 kg of germanium and an exposition of 10.3
days. We take the spectrum from Fig. 1 of Ref. \cite{cdms_lite}. We
adopt the same quenching factor that we use for CoGeNT, an energy
resolution $\sigma_{CDMSlite}=$14 eV and the efficiency
$\xi_{cut}=$0.985 \cite{cdms_lite}.

{\bf CDMS-$Ge$} We consider the data from detector T1Z5 in the range 2
keVnr$<E_R<$ 100 keVnr available in digital format from \cite{cdms_ge}
with a raw exposure of 35 kg day on germanium target. The energy
resolution is the same as in SuperCDMS, while the efficiency is taken
from Fig.1 of Ref. \cite{cdms_ge}.

{\bf SIMPLE} The SIMPLE experiment\cite{simple} uses superheated
liquid droplets homogeneously distributed in a gel to search for
transitions to the gas phase produced by WIMP scatterings. The nuclear
targets are made of $C_2 Cl F_5$ (for the nuclear response function of
chlorine, which is not available from \cite{haxton,catena}, we have
used a simple estimation outlined in Appendix
\ref{app:chlorine}). SIMPLE is a threshold detector, only sensitive to
the minimal deposited energy $E_{th}$ required to trigger the
nucleations, and with $E_{th}$ controlled by the pressure of the
liquid. The probability that an energy deposition $E_R$ on the target
nucleus $T$ nucleates a droplet is given by:

\be
P_T(E_R)=1-\exp\left [ -\alpha_T\frac{E_R-E_{th}}{E_{th}}\right ],
\label{eq:nucleation_probability}
\ee

\noindent where $\alpha_T$ is determined by fitting calibrations with
neutron sources. With an exposure of 6.71 kg day and $E_{th}$=7.8
SIMPLE observed 1 event, consistent to an expected background of
2.2. This can be converted to an upper bound of 3.16 events using the
Feldman-Cousin method \cite{feldman_cousin}. We use
$\alpha_F$=$\alpha_C$=3.6.

{\bf COUPP} The COUPP experiment\cite{coupp} searches for WIMPs using
nucleations in a bubble chamber and is also a threshold detector. In
the case of COUPP the target material is $C F_3 I$. For each operating
threshold used in COUPP the corresponding exposure, expected
background, number of measured events and 95\% C.L. upper bound
obtained with the Feldman-Cousin method \cite{feldman_cousin} used in
our analysis are summarized in Table \ref{table:coupp}. We adopt the
nucleation probability (\ref{eq:nucleation_probability}) with
$\alpha_F$=$\alpha_C$=0.15, while for iodine we assume $P_I$=1,
corresponding to $\alpha_I\rightarrow\infty$ in
(\ref{eq:nucleation_probability}).

\begin{table}[t]
\begin{center}
{\begin{tabular}{@{}|c|c|c|c|c|@{}}
\hline
$E_{th}$ (keV) & exposure (kg day) & expected background (events)  & measured events  & 95\% upper bound \\
\hline
7.8 & 55.8 &   0.8  & 2 & 5.92 \\
11 & 70 &   0.7  & 3 & 8.26 \\
15.5 & 311.7 &  3  & 8 & 12.29 \\
\hline
\end{tabular}}
\caption{For each operating threshold used in COUPP we provide the
corresponding exposure, expected background, number of measured events
and 95\% C.L. upper bound obtained with the Feldman-Cousin method
\cite{feldman_cousin} used in our analysis
  \label{table:coupp}}
\end{center}
\end{table}

{\bf PICASSO} The Picasso experiment\cite{picasso} is a bubble chamber
using $C_3 F_8$, operated with eight energy thresholds. For each of
the latter we provide the corresponding upper bound on the number of
events (normalized to events/kg/day) in Table \ref{table:picasso}
(extracted from Fig. 5 of Ref.\cite{picasso}). We use the nucleation
probability of Eq.(\ref{eq:nucleation_probability}) with
$\alpha_C$=$\alpha_F$=5.

\begin{table}[t]
\begin{center}
{\begin{tabular}{@{}|c|c|c|c|c|@{}}
\hline
$E_{th}$ (keV) & 95\% upper bound \\
\hline
1.7 &  1.1  \\
2.9  & 1.5  \\
4.1 &  11   \\
5.8  &  9   \\
6.9  & 1.3   \\
16.3  & 3.1  \\
 39 & 1.5    \\
 55  & 6     \\
\hline
\end{tabular}}
\caption{95\% C.L. upper bounds (extracted from Fig. 5 of
Ref.\cite{picasso}) for each operating threshold used in PICASSO.
  \label{table:picasso}}
\end{center}
\end{table}

{\bf PICO-2L} The PICO-2L collaboration operated a $C_3 F_8$ bubble
chamber experiment with four energy thresholds. For each of them we
provide the corresponding exposure, number of measured events and 95\%
C.L. upper bound (conservatively assuming zero background) used in our
analysis in Table \ref{table:pico2l}. In particular we conservatively
chose to use the raw data without the subtraction adopted in
\cite{pico2l} which makes use of time correlations among measured
events.  We adopt the nucleation probability
(\ref{eq:nucleation_probability}) with $\alpha_F$=$\alpha_C$=0.15.

\begin{table}[t]
\begin{center}
{\begin{tabular}{@{}|c|c|c|c|c|@{}}
\hline
$E_{th}$ (keV) & exposure (kg day) & measured events  & 95\% upper bound \\
\hline
3.2 & 74.8 &   9   & 16.77 \\
4.4 & 16.8 &   0   &  3.09 \\
6.1 &  82.2 &  3   &  8.25 \\
8.1 &  37.8 &  0   &  3.09 \\
\hline
\end{tabular}}
\caption{For each operating threshold used in PICO-2L we provide the
corresponding exposure, number of measured events
and 95\% C.L. upper bound (assuming zero background) used in our analysis.
\label{table:pico2l}}
\end{center}
\end{table}

\newpage

\section{Nuclear response functions for chlorine}
\label{app:chlorine}

In the case of chlorine a shell model calculation for the nuclear
response functions $W^{\tau\tau^{\prime}}_{\Sigma^{\prime\prime}}$ and
$W^{\tau\tau^{\prime}}_{\Sigma^{\prime}}$ is not available, so we
assume $W^{\tau\tau^{\prime}}_{\Sigma^{\prime}}= 2
W^{\tau\tau^{\prime}}_{\Sigma^{\prime\prime}}$ and use a Gaussian
approximation for the $q^2$ dependence. In particular, combining the
usual spin--dependent scaling law written as \cite{engel_spin}:

\begin{equation}
S(0)=\frac{1}{\pi}\frac{(2 j_T+1)(j_T+1)}{j_T}\left (a_p <S_p>+a_n <S_n> \right )^2
\end{equation}

\noindent with the Gaussian form factor \cite{spin_belanger}:

\begin{equation}
\frac{S(q^2)}{S(0)}=e^{-q^2 R^2/4},\,\,\,\, R=\left(0.92 A_T^{1/3}+2.68-0.78\sqrt{(A_T^{1/3}-3.8)^2+0.2} \right)\,\,\mbox{fm},
\end{equation}

\noindent (where $A_T$ is the target nucleus mass number) the identities of
Eq.(\ref{eq:spin_form_factor}) imply:

\begin{eqnarray}
  W^{\tau\tau^{\prime}}_{\Sigma^{\prime\prime}}(q^2)&=& \frac{8}{3\pi}\frac{(2 j_T+1)(j_T+1)}{j_T} <S^{\tau}><S^{\tau^{\prime}}>e^{-q^2 R^2/4} \nonumber\\
  W^{\tau\tau^{\prime}}_{\Sigma^{\prime}}(q^2)&=& \frac{4}{3\pi}\frac{(2 j_T+1)(j_T+1)}{j_T} <S^{\tau}><S^{\tau^{\prime}}>e^{-q^2 R^2/4},
\label{eq:w_approx_cl}
\end{eqnarray}
\noindent with $<S^0>=(<S_p>+<S_n>)/2$ and
$<S^1>=(<S_p>-<S_n>)/2$. For both $^{35}$Cl and $^{37}$Cl we adopt
$<S_p>$=-0.051 and $<S_p>$=-0.0088 \cite{spin_ressel}.


\begin{thebibliography}{99}

\bibitem{planck}
P.~A.~R.~Ade {\it et al.}  [Planck Collaboration],
  Astron.\ Astrophys.\  {\bf 571}, A16 (2014)
  [arXiv:1303.5076 [astro-ph.CO]].

\bibitem{dama}
 R.~Bernabei {\it et al.}  [DAMA and LIBRA Collaborations],
  Eur.\ Phys.\ J.\ C {\bf 67}, 39 (2010)
  [arXiv:1002.1028 [astro-ph.GA]].

\bibitem{lux}
D.~S.~Akerib {\it et al.}  [LUX Collaboration],
  Phys.\ Rev.\ Lett.\  {\bf 112}, no. 9, 091303 (2014)
  [arXiv:1310.8214 [astro-ph.CO]].


\bibitem{xenon100}
E.~Aprile {\it et al.}  [XENON100 Collaboration],
  Phys.\ Rev.\ Lett.\  {\bf 109}, 181301 (2012)
  [arXiv:1207.5988 [astro-ph.CO]].

\bibitem{xenon10}
J.~Angle {\it et al.}  [XENON10 Collaboration],
  Phys.\ Rev.\ Lett.\  {\bf 107}, 051301 (2011)
  [Erratum-ibid.\  {\bf 110}, 249901 (2013)]
  [arXiv:1104.3088 [astro-ph.CO]].

\bibitem{kims}
S.~C.~Kim, H.~Bhang, J.~H.~Choi, W.~G.~Kang, B.~H.~Kim, H.~J.~Kim, K.~W.~Kim and S.~K.~Kim {\it et al.},
  Phys.\ Rev.\ Lett.\  {\bf 108}, 181301 (2012)
  [arXiv:1204.2646 [astro-ph.CO]].

\bibitem{kims_modulation} Y. Kim, talk given at 13$^th$ International
  Conference on Topics in Astroparticle and Underground Physics,
  September 8--13 2013, Asilomar, California USA (TAUP2013).

\bibitem{cdms_ge}
 Z.~Ahmed {\it et al.}  [CDMS-II Collaboration],
  Phys.\ Rev.\ Lett.\  {\bf 106}, 131302 (2011)
  [arXiv:1011.2482 [astro-ph.CO]].

\bibitem{cdms_lite}
R.~Agnese {\it et al.}  [SuperCDMS Soudan Collaboration],
  Phys.\ Rev.\ Lett.\  {\bf 112}, 041302 (2014)
  [arXiv:1309.3259 [physics.ins-det]].

\bibitem{super_cdms}
R.~Agnese {\it et al.}  [SuperCDMS Collaboration],
  arXiv:1402.7137 [hep-ex].

\bibitem{simple}
M.~Felizardo, T.~A.~Girard, T.~Morlat, A.~C.~Fernandes, A.~R.~Ramos, J.~G.~Marques, A.~Kling and J.~Puibasset {\it et al.},
  Phys.\ Rev.\ Lett.\  {\bf 108}, 201302 (2012)
  [arXiv:1106.3014 [astro-ph.CO]].

\bibitem{coupp}
 E.~Behnke {\it et al.}  [COUPP Collaboration],
  Phys.\ Rev.\ D {\bf 86}, no. 5, 052001 (2012)
  [Erratum-ibid.\ D {\bf 90}, no. 7, 079902 (2014)]
  [arXiv:1204.3094 [astro-ph.CO]].

\bibitem{picasso}
S.~Archambault {\it et al.}  [PICASSO Collaboration],
  Phys.\ Lett.\ B {\bf 711}, 153 (2012)
  [arXiv:1202.1240 [hep-ex]].


\bibitem{pico2l}
C.~Amole {\it et al.}  [SNO Collaboration],
  arXiv:1503.00008 [astro-ph.CO].

\bibitem{kims_quenching_2015}
 J.~H.~Lee, G.~B.~Kim, I.~S.~Seong, B.~H.~Kim, J.~H.~Kim, J.~Li, J.~W.~Park and J.~K.~Lee {\it et al.},
  Nucl.\ Instrum.\ Meth.\ A {\bf 782}, 133 (2015)
  [arXiv:1502.03800 [physics.ins-det]].

\bibitem{factorization}
 P.~J.~Fox, J.~Liu and N.~Weiner,
  Phys.\ Rev.\ D {\bf 83}, 103514 (2011)
  [arXiv:1011.1915 [hep-ph]].

\bibitem{mccabe_eta}
C.~McCabe,
  Phys.\ Rev.\ D {\bf 84}, 043525 (2011)
  [arXiv:1107.0741 [hep-ph]];
 M.~T.~Frandsen, F.~Kahlhoefer, C.~McCabe, S.~Sarkar and K.~Schmidt-Hoberg,
  JCAP {\bf 1201}, 024 (2012)
  [arXiv:1111.0292 [hep-ph]].

\bibitem{gondolo_eta}
P.~Gondolo and G.~B.~Gelmini,
  JCAP {\bf 1212}, 015 (2012)
  [arXiv:1202.6359 [hep-ph]];
 E.~Del Nobile, G.~B.~Gelmini, P.~Gondolo and J.~H.~Huh,
  JCAP {\bf 1310}, 026 (2013)
  [arXiv:1304.6183 [hep-ph]];
  JCAP {\bf 1403}, 014 (2014)
  [arXiv:1311.4247 [hep-ph]].

\bibitem{noi1}
  S.~Scopel and K.~Yoon,
  JCAP {\bf 1408}, 060 (2014)
  [arXiv:1405.0364 [astro-ph.CO]].


\bibitem{spin_n_suppression}
  P.~Ullio, M.~Kamionkowski and P.~Vogel,
  JHEP {\bf 0107}, 044 (2001)
  [hep-ph/0010036].

\bibitem{spin_gelmini}
  E.~Del Nobile, G.~B.~Gelmini, A.~Georgescu and J.~H.~Huh,
  arXiv:1502.07682 [hep-ph].


\bibitem{spin_freitsis}
 M.~Freytsis and Z.~Ligeti,
  Phys.\ Rev.\ D {\bf 83}, 115009 (2011)
  [arXiv:1012.5317 [hep-ph]].


\bibitem{spin_arina}
  C.~Arina, E.~Del Nobile and P.~Panci,
  Phys.\ Rev.\ Lett.\  {\bf 114}, 011301 (2015)
  [arXiv:1406.5542 [hep-ph]].

\bibitem{haxton}
A.~L.~Fitzpatrick, W.~Haxton, E.~Katz, N.~Lubbers and Y.~Xu,
  JCAP {\bf 1302}, 004 (2013)
  [arXiv:1203.3542 [hep-ph]];
N.~Anand, A.~L.~Fitzpatrick and W.~C.~Haxton,
  Phys.\ Rev.\ C {\bf 89}, no. 6, 065501 (2014)
  [arXiv:1308.6288 [hep-ph]].

\bibitem{isospin_violation}
 V.~Cirigliano, M.~L.~Graesser and G.~Ovanesyan,
  JHEP {\bf 1210}, 025 (2012)
  [arXiv:1205.2695 [hep-ph]];
  V.~Cirigliano, M.~L.~Graesser, G.~Ovanesyan and I.~M.~Shoemaker,
  Phys.\ Lett.\ B {\bf 739}, 293 (2014)
  [arXiv:1311.5886 [hep-ph]].

\bibitem{noi2}
  S.~Scopel and J.~H.~Yoon,
  Phys.\ Rev.\ D {\bf 91}, no. 1, 015019 (2015)
  [arXiv:1411.3683 [hep-ph]].

\bibitem{catena}R.~Catena and B.~Schwabe,
  arXiv:1501.03729 [hep-ph].

\bibitem{engel_spin}
  J.~Engel, S.~Pittel and P.~Vogel,
  Int.\ J.\ Mod.\ Phys.\ E {\bf 1}, 1 (1992).

\bibitem{nic_ste}
N.~Fornengo and S.~Scopel,
  Phys.\ Lett.\ B {\bf 576}, 189 (2003)
  [hep-ph/0301132].

\bibitem{xenon100_response} E.~Aprile {\it et al.}  [XENON100
  Collaboration],
  Phys.\ Rev.\ D {\bf 84}, 052003 (2011)
  [arXiv:1103.0303 [hep-ex]].

\bibitem{lux_slides}
\verb!http://luxdarkmatter.org/talks/20131030_LUX_First_Results.pdf!

\bibitem{cdms_resolution}
Z.~Ahmed {\it et al.}  [CDMS Collaboration],
  Phys.\ Rev.\ D {\bf 81}, 042002 (2010)
  [arXiv:0907.1438 [astro-ph.GA]].

\bibitem{feldman_cousin}
  G.~J.~Feldman and R.~D.~Cousins,
  Phys.\ Rev.\ D {\bf 57}, 3873 (1998)
  [physics/9711021 [physics.data-an]].
  
\bibitem{spin_belanger}
G.~Belanger, F.~Boudjema, A.~Pukhov and A.~Semenov,
  Comput.\ Phys.\ Commun.\  {\bf 180}, 747 (2009)
  [arXiv:0803.2360 [hep-ph]].

\bibitem{spin_ressel}
M.~T.~Ressell, M.~B.~Aufderheide, S.~D.~Bloom, K.~Griest, G.~J.~Mathews and D.~A.~Resler,
  Phys.\ Rev.\ D {\bf 48}, 5519 (1993).

\end{thebibliography}
\end{document}